\documentclass[]{aa}
\usepackage{txfonts}
\usepackage{graphicx}
\usepackage{natbib,twoopt}
\usepackage[colorlinks,linkcolor=blue,urlcolor=blue,citecolor=black]{hyperref} 
\bibpunct{(}{)}{;}{a}{}{,}             
\makeatletter
  \newcommandtwoopt{\citeads}[3][][]{\href{http://adsabs.harvard.edu/abs/#3}%
    {\def\hyper@linkstart##1##2{}%
     \let\hyper@linkend\@empty\citealp[#1][#2]{#3}}}
  \newcommandtwoopt{\citepads}[3][][]{\href{http://adsabs.harvard.edu/abs/#3}%
    {\def\hyper@linkstart##1##2{}%
     \let\hyper@linkend\@empty\citep[#1][#2]{#3}}}
  \newcommandtwoopt{\citetads}[3][][]{\href{http://adsabs.harvard.edu/abs/#3}%
    {\def\hyper@linkstart##1##2{}%
     \let\hyper@linkend\@empty\citet[#1][#2]{#3}}}
  \newcommandtwoopt{\citeyearads}[3][][]%
    {\href{http://adsabs.harvard.edu/abs/#3}
    {\def\hyper@linkstart##1##2{}%
     \let\hyper@linkend\@empty\citeyear[#1][#2]{#3}}}
  \renewcommand*\aa@pageof{, page \thepage{} of \pageref*{LastPage}} 
\makeatother
\defcitealias{2018A&A...615L...5I}{Paper~I}
\begin{document}
\title{Hypervelocity stars in the \textit{Gaia} era} 
\subtitle{Runaway B stars beyond the velocity limit of classical ejection mechanisms}

\author{A.~Irrgang\inst{\ref{remeis}}
        \and
        S.~Kreuzer\inst{\ref{remeis}}
        \and
        U.~Heber \inst{\ref{remeis}}
}
\institute{Dr.~Karl~Remeis-Observatory \& ECAP, Astronomical Institute, Friedrich-Alexander University Erlangen-N\"urnberg (FAU), Sternwartstr.~7, 96049 Bamberg, Germany\\ \email{andreas.irrgang@fau.de}
\label{remeis}
} 

\date{Received / Accepted}

\abstract
{Young massive stars in the halo are assumed to be runaway stars from the Galactic disk. Possible ejection scenarios are binary supernova ejections (BSE) or dynamical ejections from star clusters (DE). Hypervelocity stars (HVSs) are extreme runaway stars that are potentially unbound from the Galaxy. Powerful acceleration mechanisms such as the tidal disruption of a binary system by a supermassive black hole (SMBH) are required to produce them. Therefore, HVSs are believed to originate in the Galactic center (GC), the only place known to host an SMBH.} 
{The second {\it Gaia} data release (DR2) offers the opportunity of studying HVSs in an unprecedented manner. We revisit some of the most interesting high-velocity stars, that is, 15 stars (11 candidate HVSs and 4 radial velocity outliers) for which proper motions with the {\it Hubble Space Telescope} were obtained in the pre-{\it Gaia} era, to unravel their origin.}
{By carrying out kinematic analyses based on revised spectrophotometric distances and proper motions from {\it Gaia} DR2, kinematic properties were obtained that help constrain the spatial origins of these stars.}
{Stars that were previously considered (un)bound remain (un)bound in Galactic potentials favored by {\it Gaia} DR2 astrometry. For nine stars (five candidate HVSs plus all four radial velocity outliers)
, the GC can be ruled out as spatial origin at least at $2\sigma$ confidence level, suggesting that a large portion of the known HVSs are disk runaway stars launched close to or beyond Galactic escape velocities. The fastest star in the sample, HVS\,3, is confirmed to originate in the Large Magellanic Cloud.} 
{Because the ejection velocities of five of our non-GC stars 
are close to or above the upper limits predicted for BSE and DE, another powerful dynamical ejection mechanism (e.g., involving massive perturbers such as intermediate-mass black holes)  is likely to operate in addition to the three classical scenarios mentioned above.} 

\keywords{Stars: early-type --
          Stars: kinematics and dynamics
         }
         
\maketitle

\section{Introduction}\label{sect:intro}
Hypervelocity stars (HVSs) travel so fast that they are potentially unbound from the Galaxy. When they were discovered serendipitously more than ten years ago (\citeads{2005ApJ...622L..33B}; \citeads{2005A&A...444L..61H}; \citeads{2005ApJ...634L.181E}), the so-called Hills mechanism \citepads{1988Natur.331..687H} was readily accepted as a viable ejection mechanism. According to Hills, the supermassive black hole (SMBH) at the Galactic center (GC) acts as a slingshot by tidally disrupting binary systems (see \citeads{2015ARA&A..53...15B} for a review). Because the first HVS was found to be a late B-type main-sequence (MS) star, a wide-field spectroscopic survey was initiated that targeted stars of similar spectral type and mass (2.5--4\,$M_\odot$). After surveying no less than 12\,000 square degrees of the northern sky, about two dozen unbound HVS candidates of late B-type have been discovered \citepads{2014ApJ...787...89B}. However, the Hills mechanism also ejects stars at velocities that are too low for the stars to escape the Galaxy. The respective rates are predicted to be similar to those for unbound HVSs \citepads{2009ApJ...706..925B}. Sixteen candidates for these so-called ``bound HVSs'' were identified by \citetads{2014ApJ...787...89B}. 

The HVS phenomenon is not restricted to early-type MS stars, but is also observed among evolved low-mass stars such as hot subdwarf stars (e.g., US\,708, which is also known as HVS\,2; \citeads{2005A&A...444L..61H}; \citeads{2015Sci...347.1126G}) and white dwarfs (\citeads{2017Sci...357..680V}; \citeads{2018ApJ...858....3R}; \citeads{2018arXiv180411163S}). Kinematic studies excluded the Hills scenario for those stars, suggesting that they are the surviving remnants of double-detonation supernova Ia explosions (\citeads{2015Sci...347.1126G}; \citeads{2017Sci...357..680V}). Several claims for unbound late-type stars have been rejected (\citeads{2015A&A...576L..14Z}; \citeads{2018MNRAS.479.2789B}). Hence, B-type stars constitute the main component of the known HVS population. In addition to the late B-type stars from the survey of \citetads{2014ApJ...787...89B}, even more massive and younger MS stars have been found to escape from the Galaxy. \citetads{2008A&A...483L..21H} discovered the 12\,$M_\odot$ B-giant HD\,271791 moving at hypervelocity. {\sc Hipparcos} parallaxes allowed the full six-dimensional phase space information to be exploited and the place of origin to be identified. The latter lies in the outer disk and thus excludes the Hills ejection mechanism. \citetads{2008ApJ...684L.103P} suggested that HD\,271791 is an extreme runaway B star ejected as the surviving companion of a very massive Wolf-Rayet primary that exploded in a core-collapse supernova event. The ejected star was boosted by Galactic rotation to overcome the Galactic escape velocity. HIP\,60350 is a similar candidate originating in the Galactic disk \citepads{2010ApJ...711..138I}. Finally, the set of known B-type HVSs was complemented by four stars from the Lamost spectroscopic survey (\citeads{2014ApJ...785L..23Z}; \citeads{2017ApJ...847L...9H}; \citeads{2018AJ....156...87L})

\object{HE\,0437-5439} (HVS\,3) is a particularly interesting HVS. It was discovered by \citetads{2005ApJ...634L.181E} to be a 9\,$M_\odot$ B star at a distance of 61\,kpc, which was too young (25\,Myr) to have traveled from the GC to its present position ($\approx$ 100\,Myr). \citetads{2005ApJ...634L.181E} suggested that the star was ejected from the Large Magellanic Cloud (LMC) rather than from the Galaxy because of its spatial proximity to the LMC. The LMC is not known to host an SMBH, therefore an LMC origin is controversial, and triple ejection from the GC and rejuvenation have been discussed as an alternative (\citeads{2007MNRAS.376L..29G}; \citeads{2009ApJ...698.1330P}; \citeads{2018MNRAS.475.4986F}). An LMC origin of many more HVSs was hypothesized by \citetads{2016ApJ...825L...6B} and \citetads{2017MNRAS.469.2151B}, which was motivated by the intriguing clustering of many HVSs in the constellation Leo (see \citeads{2015ARA&A..53...15B}).

Progress in HVS research can be made from high-precision astrometry. In the pre-{\it Gaia} era, the best proper motions came from the {\it Hubble Space Telescope} (HST). \citetads{2015ApJ...804...49B} determined proper motions for 12 candidate HVSs (including HVS\,2 and HVS\,3) as well as for 4 radial velocity outliers from their Hypervelocity Star Survey (see \citeads{2014ApJ...787...89B} and \citeads{2018arXiv180504184B} for the selection criteria of the survey). Already with its second data release (DR2), the {\it Gaia} mission (\citeads{2016A&A...595A...1G}; \citeads{2018A&A...616A...1G}; \citeads{2018A&A...616A...2L}) may provide proper motions superior to the HST measurements. Because the stars are very distant (30--100\,kpc), {\it Gaia} DR2 trigonometric parallaxes are of limited use. Hence, accurate and precise spectrophotometric distances are mandatory for constructing tangential velocities by combining them with {\it Gaia} proper motions. To that end, we carried out detailed quantitative spectral and photometric analyses of 14 of the late B-type stars with HST astrometry to revise their atmospheric parameters, radial and rotational velocities, masses, spectrophotometric distances, and ages (\citeads{2018A&A...615L...5I}, henceforth \citetalias{2018A&A...615L...5I}). Based on these results, we present here the kinematic analyses of that sample (including HVS\,3) by making use of {\it Gaia} DR2 proper motions.        
\section{Stellar properties}
\begin{figure}
\begin{center}
\includegraphics[width=0.49\textwidth]{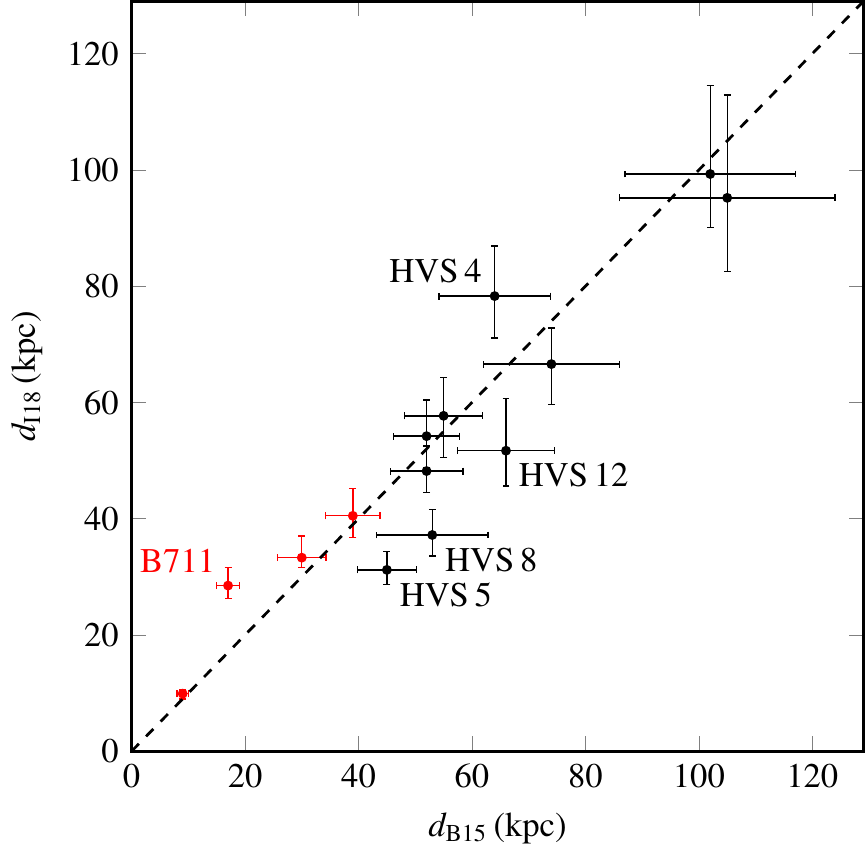}
\caption{Spectrophotometric distances $d_\textnormal{B15}$ from \citetads{2015ApJ...804...49B} are compared to the distances $d_\textnormal{I18}$ derived in \citetalias{2018A&A...615L...5I} (stars with HVS identifier are shown in black, others are red). The dashed line is the identity line. Potential outliers are labeled to facilitate identification. Error bars are $1\sigma$.}
\label{fig:dist_comp}
\end{center}
\end{figure}
\begin{figure*}
\begin{center}
\includegraphics[width=0.495\textwidth]{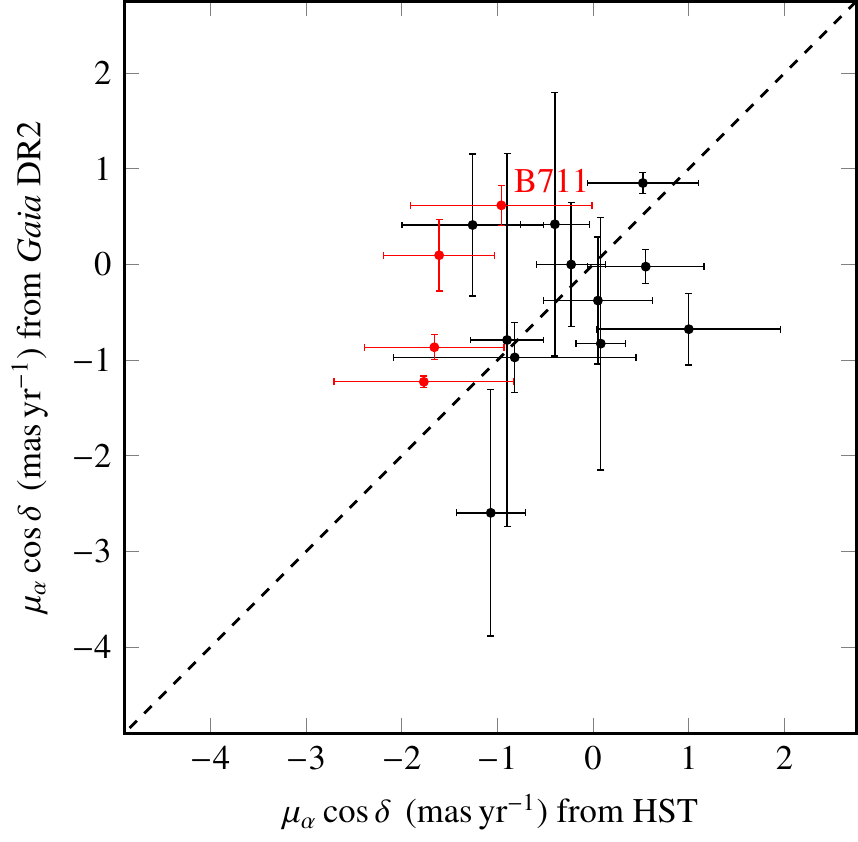}
\includegraphics[width=0.495\textwidth]{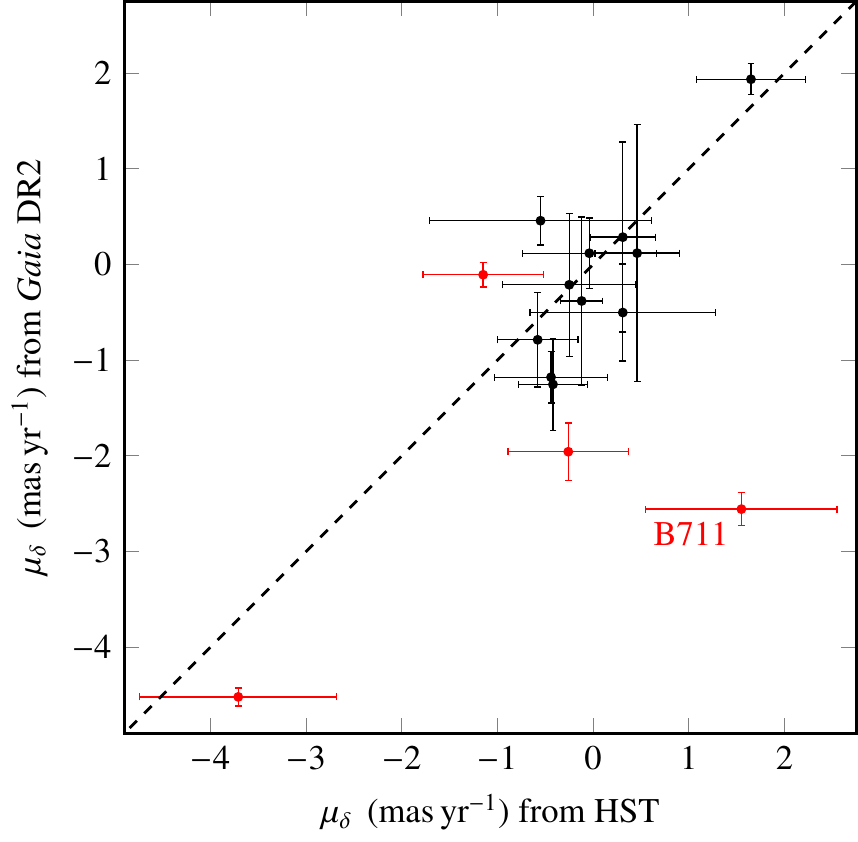}
\caption{Comparison of HST proper motions \citepads{2015ApJ...804...49B} to those from {\it Gaia} DR2 (stars with HVS identifier are shown in black, others are red). The dashed line is the identity line. Error bars are $1\sigma$.}
\label{fig:pm}
\end{center}
\end{figure*} 
\subsection{Atmospheric parameters}
Our quantitative spectral reanalysis of the MMT spectra of \citetads{2014ApJ...787...89B} confirmed that the atmospheric parameters ($T_\mathrm{eff}$, $\log(g)$) of the 14 program stars are consistent with models for MS stars with masses between 2.5\,$M_\odot$ and 5\,$M_\odot$ \citepalias{2018A&A...615L...5I}. Further support for their MS nature comes from the rotational properties of the sample. All but two stars have projected rotational velocities exceeding 50\,km\,s$^{-1}$. While being typical for MS stars, this is not expected for old horizontal branch stars of similar $T_\mathrm{eff}$ and $\log(g)$, which are slow rotators (see, e.g., \citeads{2008ASPC..392..167H}). The two exceptions (HVS\,12 and B711) exhibit projected rotational velocities that are too low ($\varv\,\sin(i)<50$\,km\,s$^{-1}$) to be precisely measured with the low-resolution MMT spectra. Radial velocities from our spectral analysis are in excellent agreement with previous determinations by \citetads{2014ApJ...787...89B}.
\subsection{Spectrophotometric distances}
\label{subsection:spectrophotometric_distances}
In addition to the spectral analysis, we investigated spectral energy distributions (SEDs) to determine stellar distances and interstellar reddening parameters. In Fig.~\ref{fig:dist_comp} we compare our spectrophotometric distance estimates to previous ones by \citetads{2015ApJ...804...49B}. For ten stars, both distance determinations are consistent. The remaining five stars, however, show partly significant discrepancies. The most severe correction ($+68$\%) was applied to B711, which we reclassified in \citetalias{2018A&A...615L...5I} as an early A-type rather than a late-B type star. The distances for HVS\,5 ($-31$\%), HVS\,8 ($-30$\%), and HVS\,12 ($-22$\%) are smaller than previously assumed, while HVS\,4 is now farther away ($+22$\%).
\subsection{Proper motions}
\label{subsection:proper_motions}
To derive the velocity vectors, proper motions are required. Prior to {\it Gaia}, the best proper motions for HVSs have been secured with HST (16 stars; \citeads{2015ApJ...804...49B}). They were obtained by measuring stellar positions with respect to galaxy reference systems in WFC3 and ACS images. It should be stressed that HST proper motions were derived from three epochs on a time base of 6.4 years from ACS and WFC3 astrometry for HVS\,1, HVS\,3, and HVS\,4, whereas only two epochs of ACS positions for HVS\,5 and two epochs of WFC3 astrometry for all other stars were available separated by about 3 years. For the brightest stars (B434, B485, B711, B733, HVS\,7, and HVS\,8), an additional systematic uncertainty arises because the positions were measured by combining shallow and deep images to measure the stellar position and to establish galaxy reference frames, respectively. As suggested by \citetads{2018arXiv180504184B}, we added 0.5\,mas\,yr$^{-1}$ in quadrature to the proper motion uncertainties given by \citetads{2015ApJ...804...49B}. We selected 15 high-velocity stars from that sample, excluding only the sdO star HVS\,2. In a first step, we compare those proper motions to {\it Gaia} DR2 values (\citeads{2018A&A...616A...2L}).

According to Fig.~\ref{fig:pm}, the HST proper motions are mostly consistent with those given by {\it Gaia} DR2, except for the inconsistent outlier B711. Like all other program stars, it satisfies the {\it Gaia} quality controls advised by \citetads{2018A&A...616A...2L}, see \citetads{2018arXiv180504184B}. We therefore studied the kinematics of B711 from both {\it Gaia} and HST proper motions. While {\it Gaia} DR2 proper motions are more precise than those of the HST for 8 of the sample stars (HVS\,3, HVS\,5, HVS\,7, HVS\,8, B434, B485, B711, and B733), this is not the case for HVS\,1, HVS\,10, HVS\,12, and HVS\,13. Although a kinematic analysis of the HST proper motions has been given in \citetads{2015ApJ...804...49B}, our revised distances also call for a reanalysis of these 4 stars. For 3 objects (HVS\,4, HVS\,6, and HVS\,9), {\it Gaia} DR2 proper motions are of similar quality as those of the HST.
   
When combined with radial velocities and spectrophotometric distances from \citetalias{2018A&A...615L...5I}, the six-dimensional phase space information is at hand (see Table~\ref{table:kinematic_short}), and kinematic analyses can be carried out.
\begin{center}
\begin{table}
\footnotesize
\setlength{\tabcolsep}{0.14cm}
\renewcommand{\arraystretch}{1.3}
\caption{\label{table:mass_models} Comparison of derived parameters of the three Milky Way mass models by \citetads{2013A&A...549A.137I} to observations.}
\begin{tabular}{lrrrr}
\hline\hline
Parameter & Model~I & Model~II & Model~III & Observation \\
\hline \hline
$M(\le 20.0\,\textnormal{kpc})$ & $2.3$ & $2.2$ & $2.7$ & $1.91^{+0.17}_{-0.15}$\tablefootmark{a} \\
$M(\le 39.5\,\textnormal{kpc})$ & $4.1$ & $3.8$ & $6.1$ & $4.4^{+0.7}_{-0.6}$\tablefootmark{b}; $6.1^{+1.8}_{-1.2}$\tablefootmark{c} \\
$M(\le 200\,\textnormal{kpc})$  & $19$  & $12$  & $30$ & $M_\mathrm{MW} \ge 9.1^{+6.2}_{-2.6}$\tablefootmark{d} \\
$\varv_{\mathrm{esc},\odot}$ & $616$ & $576$ & $812$ & $\sim 600$\tablefootmark{e}; $580 \pm 63$\tablefootmark{f} \\
\hline
\end{tabular}
\tablefoot{The quantity $M(\le R)$ is the total mass enclosed within a radius $R$ and is given in units of $10^{11}M_\odot$. The Galactic escape velocity at the Sun's position, $\varv_{\mathrm{esc},\odot}$, is in km\,s$^{-1}$.}
\tablebib{
\tablefoottext{a}{\citetads{2018arXiv180501408P}};
\tablefoottext{b}{\citetads{2018arXiv180411348W}};
\tablefoottext{c}{\citetads{2018ApJ...862...52S}};
\tablefoottext{d}{\citetads{2018A&A...616A..12G}};
\tablefoottext{e}{\citetads{2018arXiv180503194H}};
\tablefoottext{f}{\citetads{2018A&A...616L...9M}}.
}
\end{table}
\end{center}
\begin{center}
\begin{table*}
\footnotesize
\setlength{\tabcolsep}{0.1475cm}
\renewcommand{\arraystretch}{1.3}
\caption{\label{table:kinematic_short}Input and output parameters of the kinematic analyses.}
\begin{tabular}{lrrrrrrrrrrrrrrrrrrrrr}
\hline\hline
Object & \multicolumn{2}{c}{$d$} && \multicolumn{1}{c}{$\mu_\alpha \cos\delta$} & \multicolumn{1}{c}{$\mu_\delta$} && Corr. && \multicolumn{2}{c}{$\varv_\mathrm{rad}$} & \multicolumn{2}{c}{$\varv_\mathrm{Grf}$} & \multicolumn{2}{c}{$\varv_\mathrm{ej,p}$} && $P_\mathrm{b}$ && \multicolumn{2}{c}{$\tau_\mathrm{flight,p}$} & \multicolumn{2}{c}{$\tau$} \\
\cline{2-3} \cline{5-6} \cline{10-15} \cline{19-22} 
& \multicolumn{2}{c}{(kpc)} && \multicolumn{2}{c}{(mas\,yr$^{-1}$)} &&&& \multicolumn{6}{c}{(km\,s$^{-1}$)} && (\%) && \multicolumn{4}{c}{(Myr)}\\
\hline\hline
 HVS\,1 (H) & $99.3$ & $^{+15.2}_{-\phantom{0}9.2}$ && $ 0.08 \pm 0.26$ & $-0.12 \pm 0.22$ &&  \ldots && $829.7$ & $^{+2.2 }_{-2.2 }$ & $690$ & $^{+40}_{-20}$             & $750$  & $^{+90}_{-90}$             &&   $0$ && $112$ & $^{+34}_{-22}$           & $272$ & $^{+12}_{-11}$ \\
     HVS\,3 & $62.3$ & $^{+7.7 }_{-7.7 }$           && $ 0.85 \pm 0.11$ & $ 1.94 \pm 0.17$ &&  $0.19$ && $723.0$ & $^{+1.2 }_{-1.2 }$ & $820$ & $^{+70}_{-70}$             & \ldots & \ldots                     &&   $0$ && $104$ & $^{+20}_{-17}$           &  $18$ & $^{+3}_{-3}$ \\
     HVS\,4 & $78.3$ & $^{+8.6 }_{-7.2 }$           && $ 0.00 \pm 0.65$ & $-1.25 \pm 0.49$ && $-0.67$ && $604.6$ & $^{+2.8 }_{-2.6 }$ & $630$ & $^{+120}_{-\phantom{0}60}$ & $840$  & $^{+\phantom{0}70}_{-130}$ &&   $0$ && $129$ & $^{+56}_{-35}$           & $150$ & $^{+\phantom{0}6}_{-10}$ \\
     HVS\,5 & $31.2$ & $^{+3.2 }_{-2.5 }$           && $-0.02 \pm 0.18$ & $-1.18 \pm 0.27$ &&  $0.22$ && $542.5$ & $^{+2.9 }_{-3.0 }$ & $650$ & $^{+10}_{-10}$             & $640$  & $^{+50}_{-40}$             &&   $0$ &&  $46$ & $^{+4}_{-5}$             &  $97$ & $^{+31}_{-37}$ \\
     HVS\,6 & $57.7$ & $^{+6.6 }_{-7.2 }$           && $-0.38 \pm 0.67$ & $-0.50 \pm 0.51$ && $-0.02$ && $619.3$ & $^{+3.8 }_{-4.1 }$ & $550$ & $^{+60}_{-30}$             & $680$  & $^{+90}_{-80}$             &&   $0$ &&  $96$ & $^{+21}_{-17}$           & $142$ & $^{+33}_{-54}$ \\
     HVS\,7 & $48.2$ & $^{+4.3 }_{-3.7 }$           && $-0.68 \pm 0.38$ & $ 0.46 \pm 0.26$ && $-0.28$ && $524.0$ & $^{+1.7 }_{-1.5 }$ & $500$ & $^{+50}_{-40}$             & $530$  & $^{+30}_{-30}$             &&   $5$ &&  $82$ & $^{+10}_{-\phantom{0}8}$ & $185$ & $^{+\phantom{0}7}_{-10}$ \\
     HVS\,8 & $37.2$ & $^{+4.4 }_{-3.6 }$           && $-0.97 \pm 0.37$ & $ 0.12 \pm 0.37$ && $-0.38$ && $499.6$ & $^{+3.4 }_{-3.4 }$ & $500$ & $^{+50}_{-40}$             & $450$  & $^{+40}_{-30}$             &&  $16$ &&  $87$ & $^{+18}_{-14}$           & $226$ & $^{+24}_{-51}$ \\
     HVS\,9 & $66.6$ & $^{+6.2 }_{-7.0 }$           && $ 0.41 \pm 0.75$ & $-0.21 \pm 0.75$ && $-0.25$ && $622.0$ & $^{+3.1 }_{-3.0 }$ & $570$ & $^{+140}_{-\phantom{0}80}$ & $690$  & $^{+110}_{-120}$           &&   $0$ &&  $90$ & $^{+34}_{-21}$           & $175$ & $^{+\phantom{0}8}_{-24}$ \\
HVS\,10 (H) & $54.2$ & $^{+6.2 }_{-5.4 }$           && $-1.07 \pm 0.36$ & $-0.58 \pm 0.42$ &&  \ldots && $462.0$ & $^{+5.0 }_{-2.2 }$ & $450$ & $^{+60}_{-30}$             & $600$  & $^{+70}_{-40}$             &&  $29$ && $114$ & $^{+18}_{-14}$           & $210$ & $^{+50}_{-80}$ \\
HVS\,12 (H) & $51.7$ & $^{+9.0 }_{-6.1 }$           && $ 0.40 \pm 0.36$ & $ 0.31 \pm 0.34$ &&  \ldots && $545.0$ & $^{+3.4 }_{-3.4 }$ & $500$ & $^{+60}_{-50}$             & $510$  & $^{+40}_{-30}$             &&   $8$ &&  $88$ & $^{+19}_{-14}$           &  $90$ & $^{+77}_{-34}$ \\
HVS\,13 (H) & $95.2$ & $^{+17.7}_{-12.7}$           && $-0.90 \pm 0.38$ & $ 0.46 \pm 0.44$ &&  \ldots && $568.1$ & $^{+5.1 }_{-5.5 }$ & $690$ & $^{+170}_{-150}$           & $610$  & $^{+170}_{-100}$           &&   $0$ && $179$ & $^{+72}_{-44}$           & $200$ & $^{+32}_{-77}$ \\
       B434 & $40.5$ & $^{+4.7 }_{-3.7 }$           && $ 0.10 \pm 0.38$ & $-1.96 \pm 0.30$ && $-0.10$ && $445.5$ & $^{+2.5 }_{-2.3 }$ & $380$ & $^{+50}_{-40}$             & $590$  & $^{+20}_{-20}$             &&  $92$ && $118$ & $^{+26}_{-19}$           & $402$ & $^{+16}_{-23}$ \\
       B485 & $33.3$ & $^{+3.7 }_{-1.7 }$           && $-0.87 \pm 0.14$ & $-0.11 \pm 0.13$ && $-0.51$ && $422.9$ & $^{+1.8 }_{-1.3 }$ & $450$ & $^{+20}_{-20}$             & $420$  & $^{+20}_{-10}$             &&  $89$ &&  $83$ & $^{+11}_{-\phantom{0}6}$ &  $94$ & $^{+5}_{-5}$ \\
   B711     & $28.5$ & $^{+3.1 }_{-2.2 }$           && $ 0.62 \pm 0.21$ & $-2.56 \pm 0.18$ && $ 0.02$ && $271.0$ & $^{+1.2 }_{-1.4 }$ & $420$ & $^{+30}_{-30}$             & $440$  & $^{+10}_{-10}$             &&  $99$ && $113$ & $^{+22}_{-14}$           & $393$ & $^{+59}_{-17}$ \\
   B711 (H) & $28.5$ & $^{+3.1 }_{-2.2 }$           && $-0.96 \pm 0.95$ & $ 1.55 \pm 1.00$ &&  \ldots && $271.0$ & $^{+1.2 }_{-1.4 }$ & $510$ & $^{+120}_{-110}$           & $600$  & $^{+90}_{-50}$             &&  $48$ &&  $61$ & $^{+11}_{-\phantom{0}8}$ & $393$ & $^{+59}_{-17}$ \\
       B733 &  $9.9$ & $^{+0.7 }_{-0.9 }$           && $-1.23 \pm 0.06$ & $-4.52 \pm 0.10$ &&  $0.69$ && $350.8$ & $^{+1.6 }_{-1.4 }$ & $460$ & $^{+10}_{-10}$             & $450$  & $^{+10}_{-10}$             && $100$ &&  $22$ & $^{+2}_{-2}$             & $123$ & $^{+40}_{-47}$ \\
\hline
\end{tabular}
\tablefoot{Spectrophotometric distances $d$, radial velocities $\varv_{\mathrm{rad}}$, and stellar ages $\tau$ are from \citetalias{2018A&A...615L...5I} (except for HVS\,3, whose age and radial velocity are from \citeads{2008A&A...480L..37P}). Proper motions $\mu_\alpha \cos\delta$, $\mu_\delta$, and their correlations (\textit{``Corr.'' column}) are from {\it Gaia} DR2. A suffix ``(H)'' in the \textit{``Object'' column} indicates that proper motions are from the HST. The quantities $\varv_{\mathrm{Grf}}$ (current Galactic rest-frame velocity), $\varv_{\mathrm{ej,p}}$ (ejection velocity from the plane corrected for Galactic rotation), $P_{\mathrm{b}}$ (probability to be bound to the Galaxy), and $\tau_{\mathrm{flight,p}}$ (flight time from the Galactic plane) result from trajectories calculated in Model~I of \citetads{2013A&A...549A.137I}. More information on the trajectories is listed in Table \ref{table:kinematic_parameters_I} and visualized in Figs.~\ref{fig:orbits} and \ref{fig:orbits_appendix}. The given uncertainties are $1\sigma$.}
\end{table*}
\end{center}
\section{Galactic mass models and stellar trajectories}
\label{sect:kinematic}
To trace back stellar trajectories to the Galactic plane, we numerically integrated the equations of motion resulting from three different Milky Way mass models (see \citeads{2013A&A...549A.137I} for details). Model~I is based on the potential of \citetads{1991RMxAA..22..255A}, Model~II is the truncated, flat rotation curve model of \citetads{1999MNRAS.310..645W}, and Model~III is the widely used one by \citetads{1997ApJ...490..493N} derived from numerical cosmological simulations. \citetads{2013A&A...549A.137I} updated the model parameters of all three potentials by making use of observational constraints from the Galactic rotation curve, the proper motion of Sgr\,A$^\ast$, the local mass surface density, the velocity dispersion in Baade's window, and the assumption that the kinematically hottest halo star is bound. Recent analyses based on {\it Gaia} DR2 astrometry of the motions of globular clusters, satellite galaxies, and extreme velocity halo stars seem to favor Models~I and~II over Model~III (see Table~\ref{table:mass_models}). In particular, the huge escape velocity from the solar neighborhood predicted by Model~III is at odds with the observational results.

Monte Carlo simulations were carried out to propagate the uncertainties of the input parameters (spectrophotometric distance, radial velocity, and proper motions) assuming Gaussian distributions for each parameter while also accounting for asymmetric error bars and the correlation between the two proper motion components. The difference between Galactic rest-frame and the local escape velocity determines whether a star is bound ($\varv_{\mathrm{Grf}}-\varv_{\mathrm{esc}}<0$) to the Galaxy or not ($\varv_{\mathrm{Grf}}-\varv_{\mathrm{esc}}>0$). The fraction of Monte Carlo runs for which the star is bound to the Milky Way is denoted as the probability $P_{\mathrm{b}}$. The trajectories were integrated backward in time for twice the stellar age, except for HVS\,3, whose trajectories were calculated for 200\,Myr (see Sect.~\ref{sect:hvs3_lmc}). The corresponding Galactic plane-crossing quantities allow the stars' place of origin to be constrained. The ejection velocities $\varv_{\mathrm{ej}}$ (defined as the Galactic rest-frame velocity relative to the rotating Galactic disk) impose strong constraints on the ejection mechanisms. Moreover, the time of flight from the plane to the present location of the star must not exceed its age. All relevant kinematic quantities derived from the three different mass models are listed in the appendix (Tables~\ref{table:kinematic_parameters_I}, \ref{table:kinematic_parameters_II}, and \ref{table:kinematic_parameters_III}). As expected for very fast objects, the Galactic plane-crossing quantities are almost independent of the choice of the Galactic mass model. In contrast, the bound probability is very model sensitive because the adopted mass of the dark matter halo differs. For the lowest Milky Way mass (Model~II), HVS\,1 to HVS\,13 have zero probability to be bound (see Table~\ref{table:kinematic_parameters_II}), while for the most massive halo (Model~III), only HVS\,1 has zero probability to be bound, the probabilities for HVS\,3, HVS\,4, and HVS\,13 are lower than 50\%, for HVS\,6, HVS\,9, and HVS\,12 they are higher than 50\%, and HVS\,5, HVS\,7, HVS\,8, and HVS\,10 have 100\% probability to be bound (see Table~\ref{table:kinematic_parameters_III}). The four velocity outliers (B434, B485, B711, and B733) are very likely bound in Models~I and III, while B485 would be unbound in Model~II.

\begin{figure*}
\begin{center}
\includegraphics[width=0.33\textwidth]{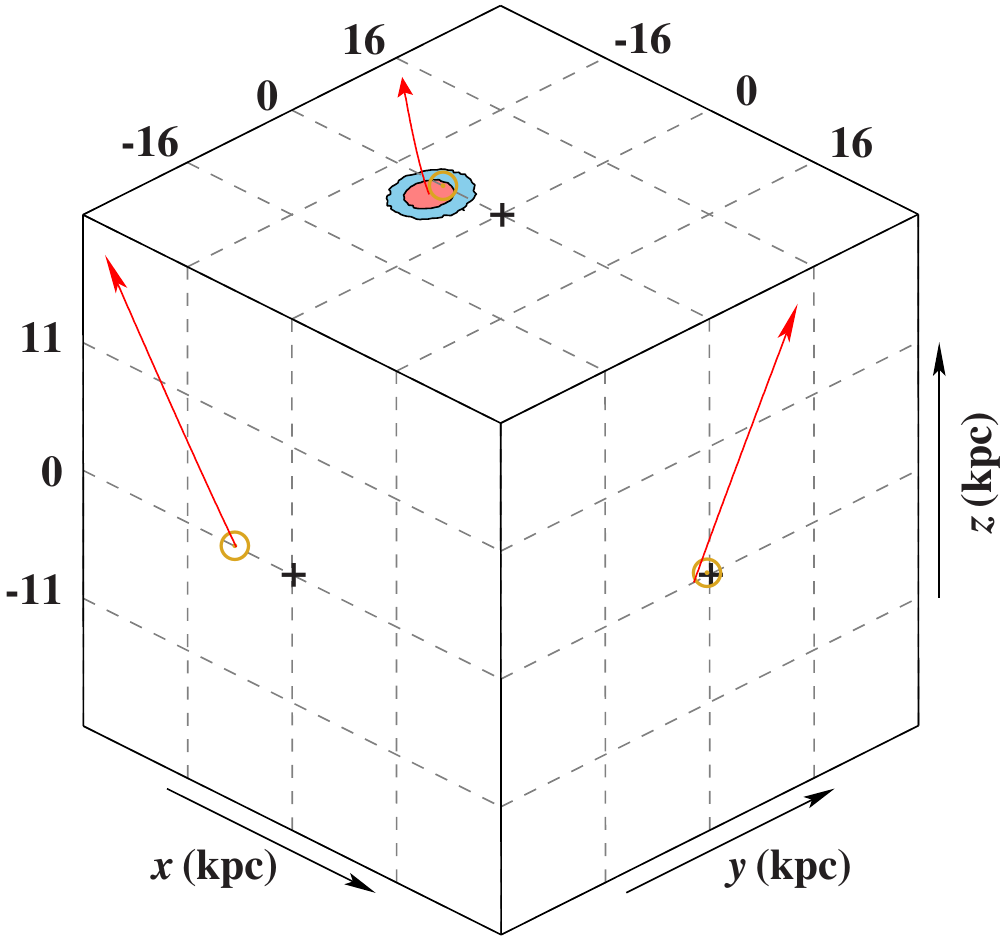}
\includegraphics[width=0.33\textwidth]{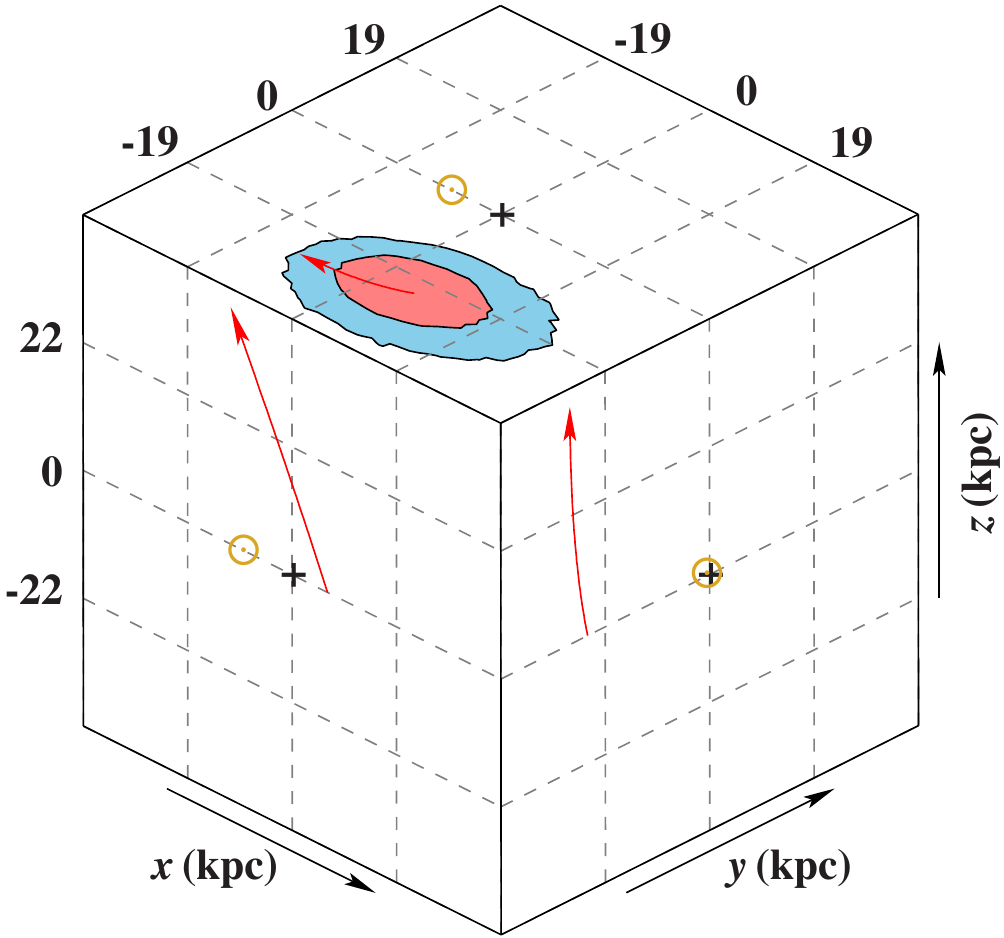}
\includegraphics[width=0.33\textwidth]{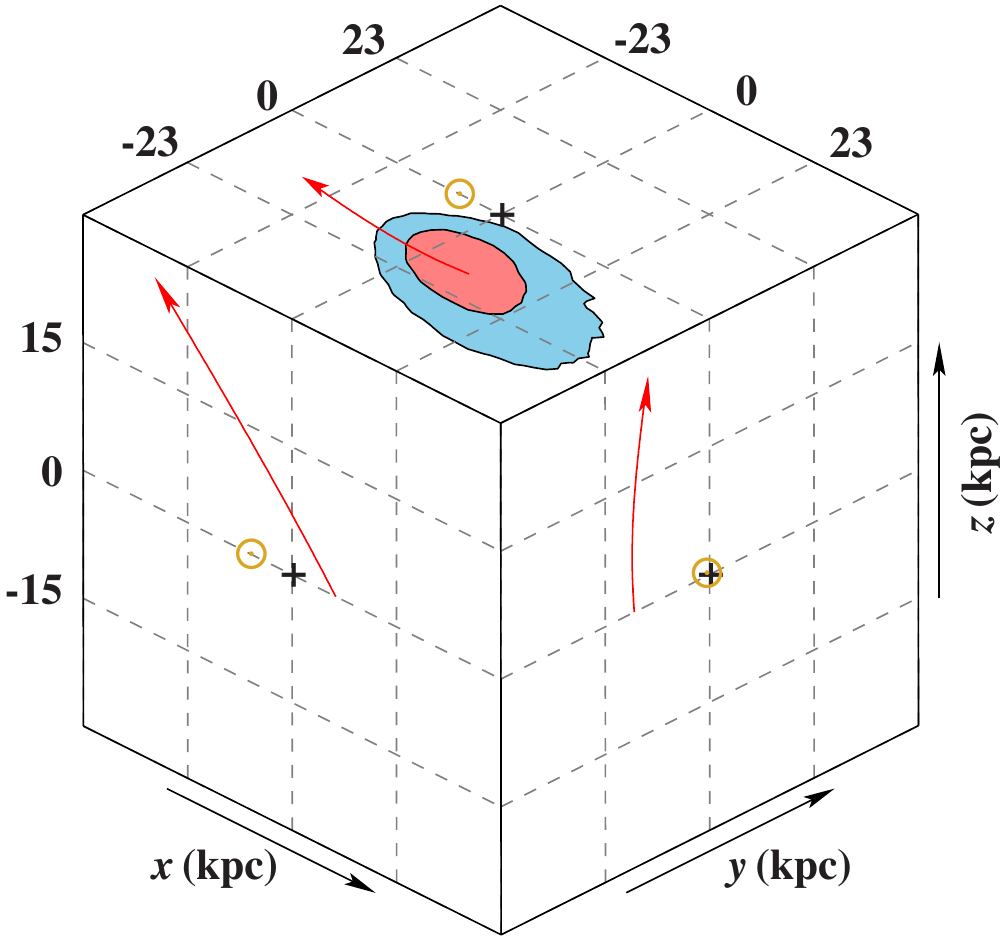}
\includegraphics[width=0.33\textwidth]{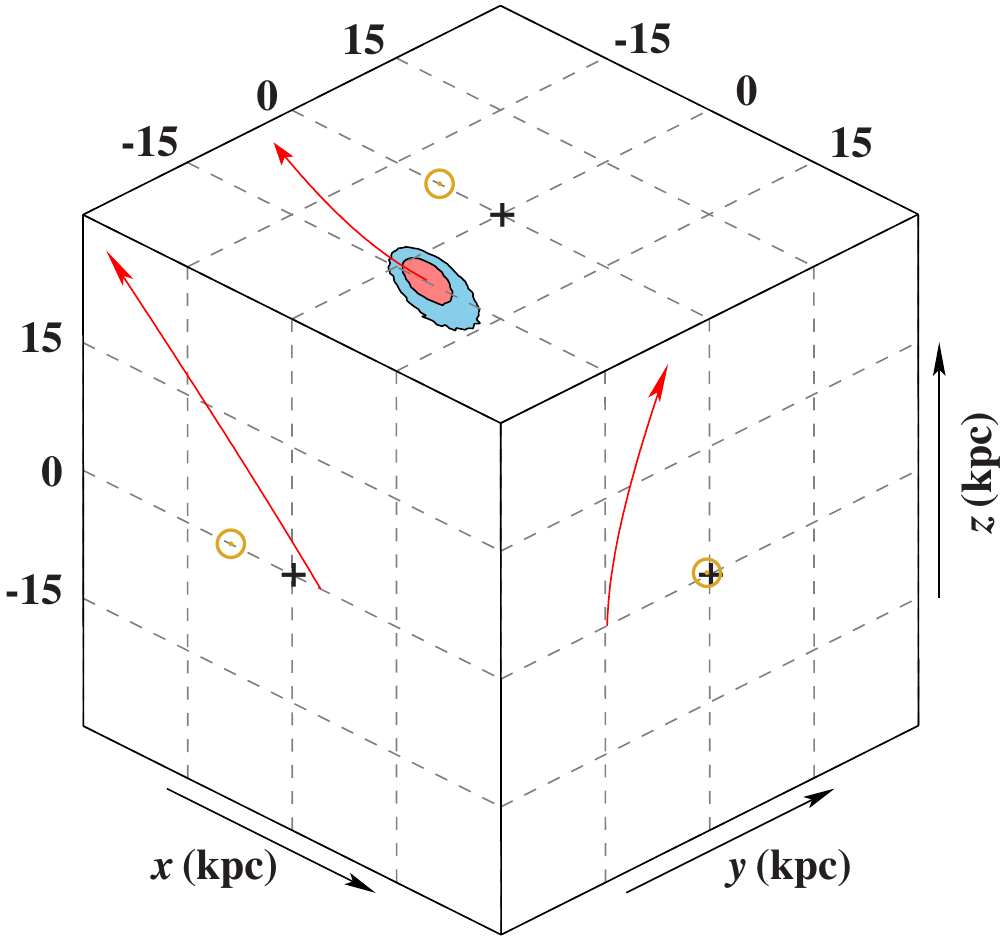}
\includegraphics[width=0.33\textwidth]{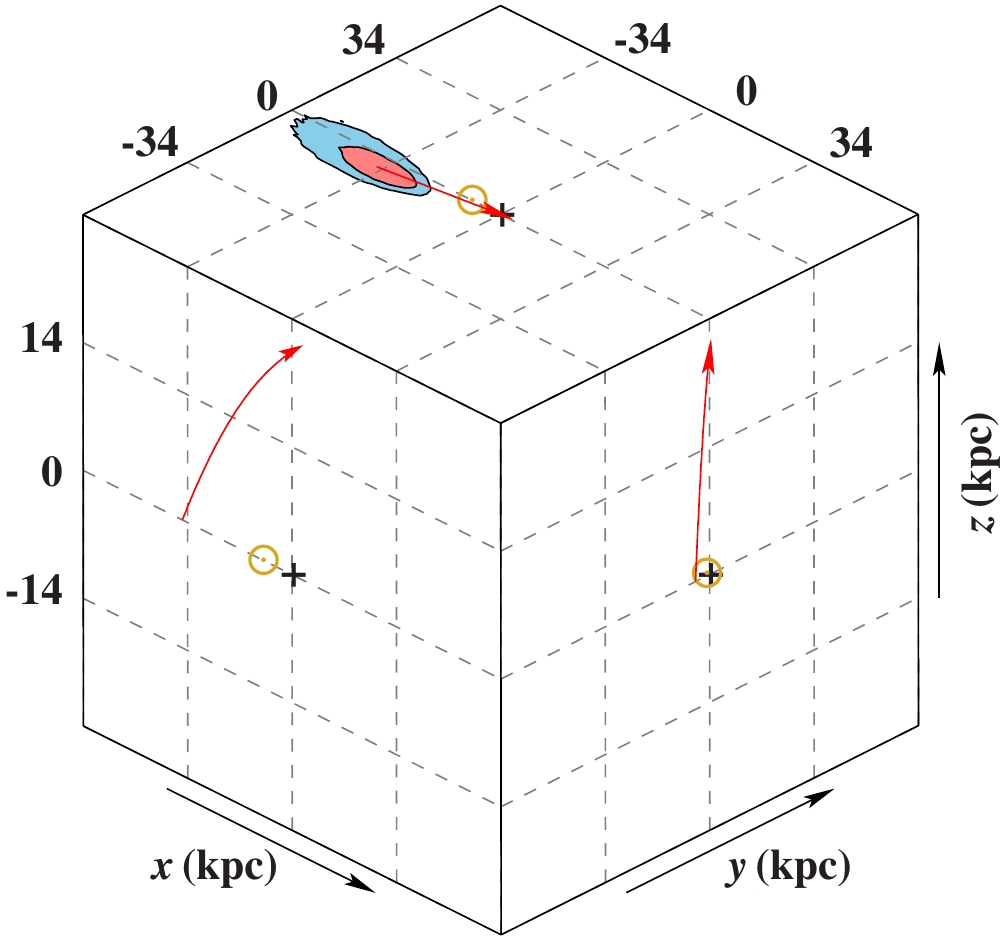}
\includegraphics[width=0.33\textwidth]{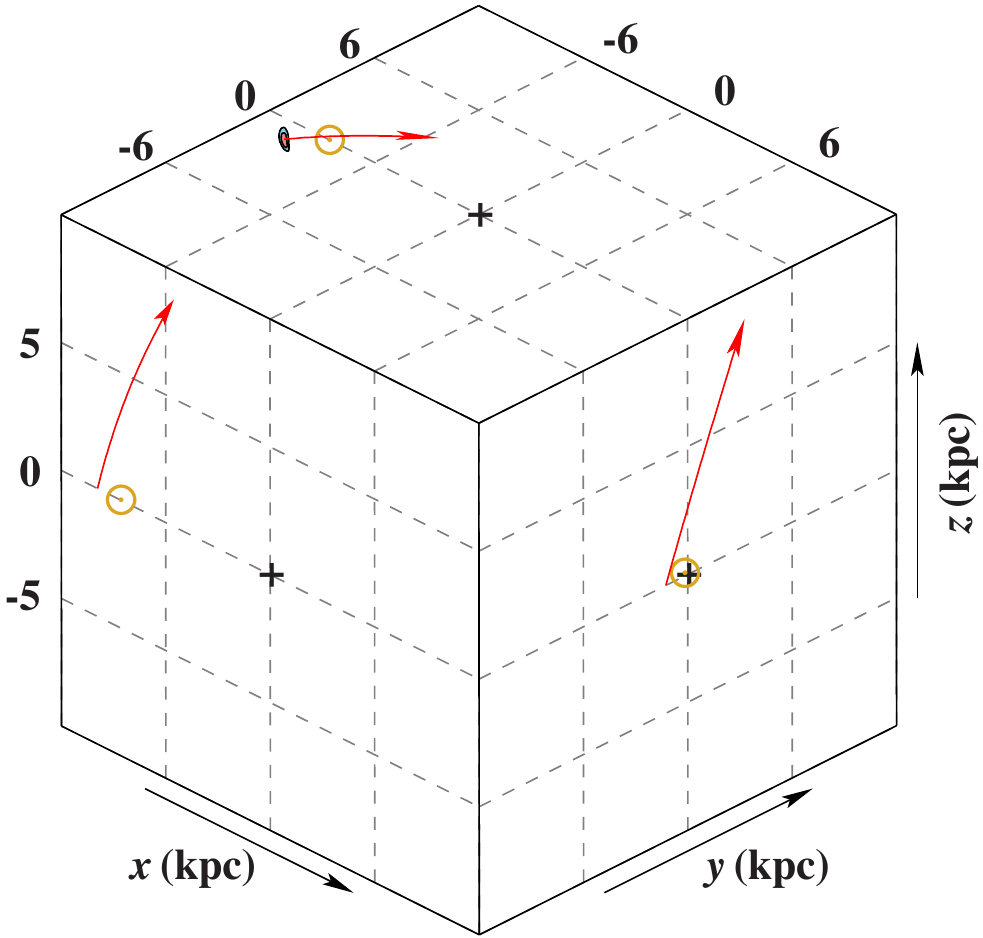}
\caption{Three-dimensional orbits of HVS\,5 (\textit{top left}), HVS\,7 (\textit{top middle}), HVS\,8 (\textit{top right}), B485 (\textit{bottom left}), B711 (\textit{bottom middle}), and B733 (bottom right) in a Galactic Cartesian coordinate system in which the $z$-axis points to the Galactic north pole. The trajectories (red line; the arrow indicates the current position of the star) are traced back to the Galactic plane using Model~I of \citetads{2013A&A...549A.137I}. The black rimmed, red and blue shaded areas mark regions where 68\% and 95\% ($1\sigma$ and $2\sigma$) of the trajectories intersected the Galactic plane when uncertainties in the distance, proper motions, and radial velocity were propagated. Orbits that did not cross the Galactic plane within twice the estimated stellar lifetime were omitted to account for the finite age of the star. The positions of the Sun and the GC are marked by a yellow $\odot$ and a black $+$, respectively.
\label{fig:orbits}}
\end{center}
\end{figure*}

In Table~\ref{table:kinematic_short} we summarize the most important results based on Model~I, that is, the current Galactic rest-frame velocity, ejection velocity, probability for a star to be bound, and the time of flight to reach the Galactic plane. These quantities are complemented by the spectrophotometric distance, radial velocity, and stellar age from \citetalias{2018A&A...615L...5I}. The three-dimensional trajectories are shown in Figs.~\ref{fig:orbits} and \ref{fig:orbits_appendix}. 
\subsection{Plane-crossing properties}
\begin{figure*}
\begin{center}
\includegraphics[width=0.99\textwidth]{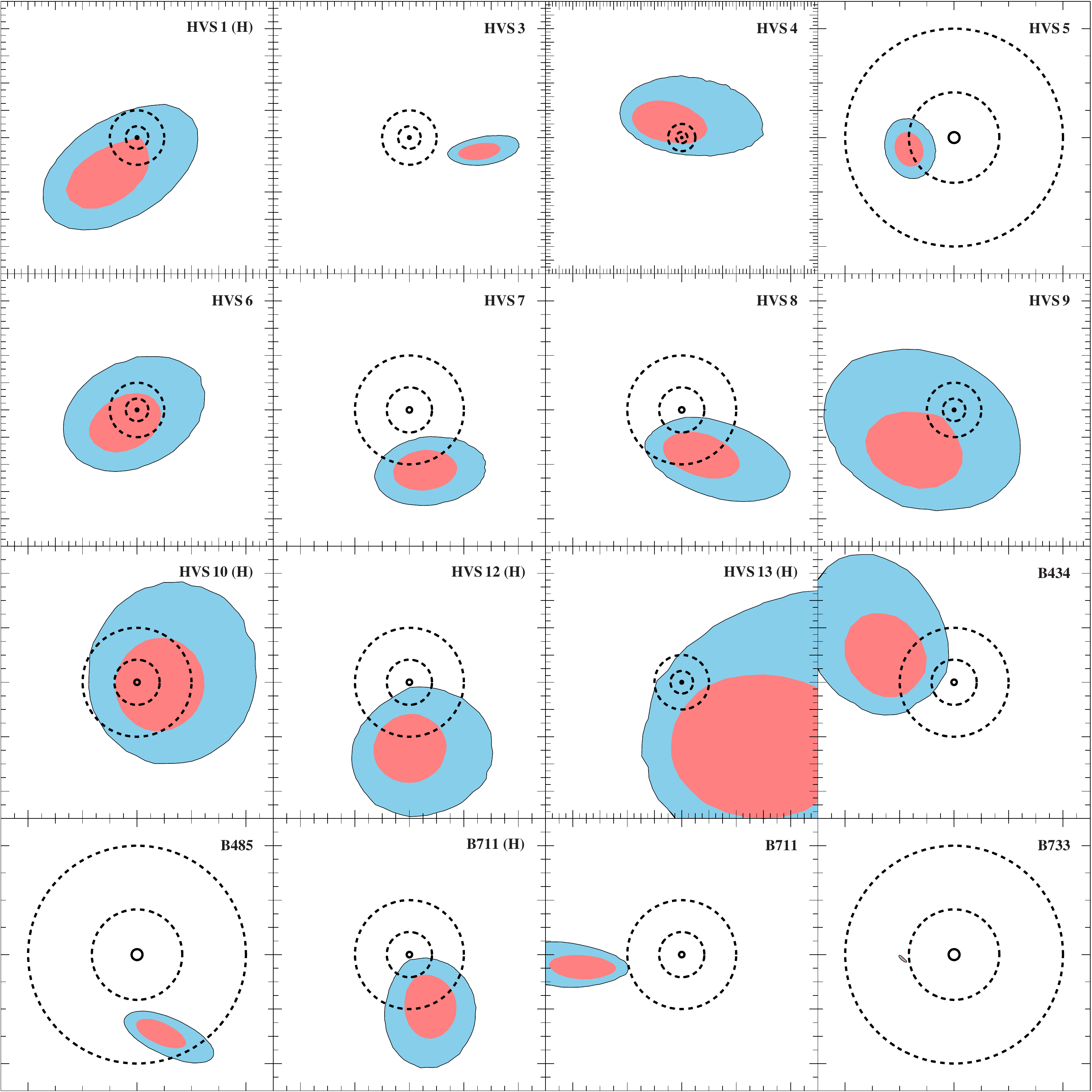}
\caption{Galactic plane-crossing locations of the program stars (the black rimmed, red and blue shaded areas have the same meaning as in Fig.~\ref{fig:orbits}). The full circle in the middle indicates the position of the GC. The dashed circles have radii of 8 and 20\,kpc, thus roughly indicating the solar circle and the outer edge of the Galactic disk, which rotates clockwise in this plot. Because {\it Gaia} DR2 and HST proper motions are inconsistent for B711, the trajectories were calculated for both cases. Results based on proper motions from HST are marked by the suffix (H).}
\label{fig:grid}
\end{center}
\end{figure*}
The plane-crossing properties are listed in Tables~\ref{table:kinematic_parameters_I} to \ref{table:kinematic_parameters_III} and plotted in Fig.~\ref{fig:grid}. Because they are basically independent of the choice of the Galactic potential, we restrict ourselves to discuss the properties derived from Model~I (Table~\ref{table:kinematic_parameters_I}). 
\subsubsection{HVSs}
HVS\,3: As shown in Fig.~\ref{fig:grid}, the star does not cross the Galactic plane anywhere near the GC or disk. The plane-crossing contours lie well beyond the 20\,kpc radius adopted here as estimate for the size of the Galactic disk. Its time of flight ($104^{+20}_{-17}$\,Myr) is more than five times its evolutionary lifetime ($18\pm3$\,Myr), confirming the results of previous studies (\citeads{2005ApJ...634L.181E}; \citeads{2008A&A...480L..37P}) and providing additional evidence against an origin in the Milky Way. Because HVS\,3 has been suggested to originate from the LMC, we investigate this option in detail in Sect.~\ref{sect:hvs3_lmc}.

HVS\,1, HVS\,4, HVS\,6, HVS\,9, HVS\,10, and HVS\,13: Available astrometry is insufficient to narrow down the place of origin other than to the full Galactic disk (see Fig.~\ref{fig:grid}), thus also covering the GC. This is consistent with the HST-based result (see Fig.~4 in \citeads{2015ApJ...804...49B}). Three-dimensional trajectories are shown Fig.~\ref{fig:orbits_appendix}. We have to await the next {\it Gaia} data releases to draw further conclusions.

HVS\,5, HVS\,7, HVS\,8, and HVS\,12: {\it Gaia} DR2 astrometry and/or our revised distances are sufficient to exclude the GC at a confidence level of more than 2$\sigma$ (see Figs.~\ref{fig:orbits}, \ref{fig:orbits_appendix}, and \ref{fig:grid}). HVS\,5 originates in the disk close to the solar circle, while it is likely that HVS\,7, HVS\,8, and HVS\,12 come from the outer rims of the Galactic disk. All four stars are therefore disk-runaway stars. However, their high ejection velocities, which range from $450^{+40}_{-30}$\,km\,s$^{-1}$ (HVS\,8) to $640^{+50}_{-40}$\,km\,s$^{-1}$ (HVS\,5), are a challenge to the ejection scenarios (see Sect.~\ref{sect:discussion}). In addition to HVS\,10, HVS\,8 and HVS\,12 have the highest probabilities to be bound to the Galaxy among the HVSs (see Table~\ref{table:kinematic_parameters_I}). A higher Galactic halo mass than anticipated in Model~I would render them bound to the Galaxy.
\subsubsection{Radial velocity outliers} 
The plane-crossing properties of B434, B485, B711, and B733 clearly show that these objects are runaway stars from the Galactic disk rather than bound HVSs from the GC. Their ejection velocities, which range from $420^{+20}_{-10}$\,km\,s$^{-1}$ for B485 to $590^{+20}_{-20}$\,km\,s$^{-1}$ for B434, are almost as high as that of the presumably unbound disk-runaway stars HVS\,5, HVS\,7, HVS\,8, and HVS\,12.

B733: This object is much brighter ($g=15.67$\,mag) and closer ($9.9$\,kpc) than any other sample star. Therefore, the {\it Gaia} DR2 proper motions are of excellent quality and allow its place of origin in the Galactic disk to be pinned down to a narrow region close to the solar circle (see Figs.~\ref{fig:orbits} and \ref{fig:grid}).

B485: This is probably the youngest ($94^{+5}_{-5}$\,Myr), most massive ($4.8\,M_\odot$), and hottest ($T_\mathrm{eff}=15\,200^{+370}_{-410}$\,K) sample star after HVS\,3. Its place of origin is in the outer disk, but it is not yet constrained as precisely as that of B733 (see Figs.~\ref{fig:orbits} and \ref{fig:grid}).

B434: This star is the oldest ($402^{+16}_{-23}$\,Myr) in the sample. Its plane-crossing properties are still not well defined (see Figs.~\ref{fig:grid} and \ref{fig:orbits_appendix}). Nevertheless, the GC can be excluded at more than 2$\sigma$ confidence, and its origin appears to be at the outer rim of the Galactic disk.

B711: This object is the coolest ($T_\mathrm{eff}=9170^{+230}_{-250}$\,K) and most evolved in the sample. Owing to its low effective temperature, we have reclassified B711 as an A-type star in \citetalias{2018A&A...615L...5I}. Because the {\it Gaia} and HST proper motions are discrepant (see Sect.~\ref{subsection:proper_motions}), we performed the kinematic analysis of B711 twice. As expected, the plane-crossing properties differ widely (see Fig.~\ref{fig:grid}). The {\it Gaia} DR2 astrometry would place it outside of the 20\,kpc circle, while the HST one would allow for an origin in the Galactic disk. The latter appears to be more plausible but implies a very high ejection velocity of $600^{+90}_{-50}$\,km\,s$^{-1}$.   
\subsection{Time of flight versus stellar age}
\begin{figure}
\begin{center}
\includegraphics[width=0.477\textwidth]{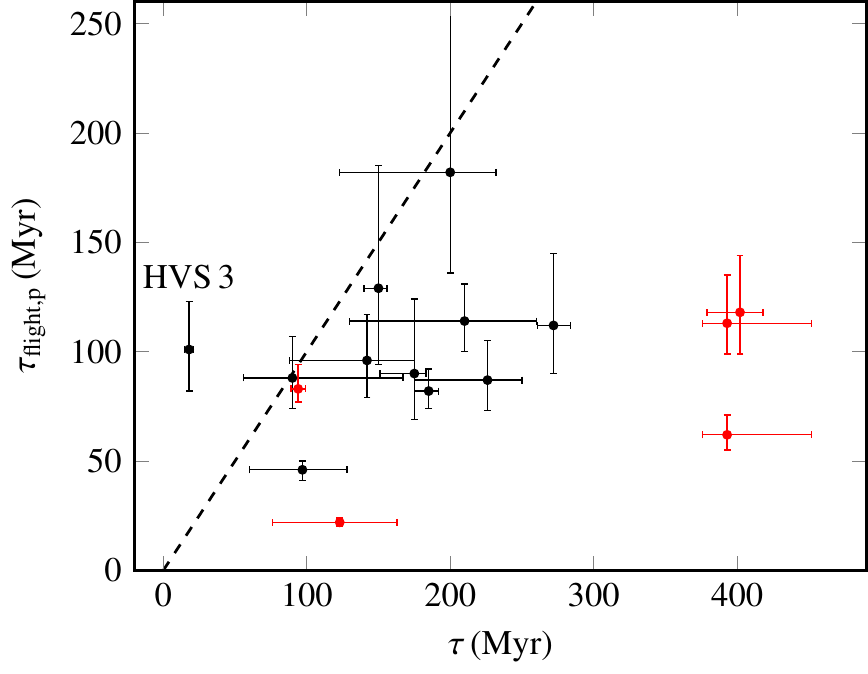}
\caption{Flight times $\tau_\textnormal{flight,p}$ (to reach the Galactic plane) vs.\ stellar age $\tau$ (stars with HVS identifier are shown in black, others are red). The dashed line is the identity line. Error bars are $1\sigma$.}
\label{fig:tof_age}
\end{center}
\end{figure}
If the assumption of an origin in the Galactic disk is correct, the times of flight from the Galactic plane must be shorter than the stellar ages. In Fig.~\ref{fig:tof_age} these two quantities are plotted against each other. The times of flight are indeed shorter than the stellar lifetimes for all program stars except for HVS\,3, which we discuss in the following section.
\section{LMC origin of HVS 3}\label{sect:hvs3_lmc}
Because of its proximity to the LMC on the sky ($\approx$16$^\circ$), \citetads{2005ApJ...634L.181E} suggested that HVS\,3 originated in the LMC. A differential abundance analysis of high-resolution spectra (\citeads{2008A&A...480L..37P}; \citeads{2008ApJ...675L..77B}) strengthened this idea because a subsolar abundance pattern was found, consistent with patterns of B-type stars in the LMC. Proper motions from the HST allowed reconstructing the trajectory of HVS\,3, which led to inconclusive results, however (\citeads{2013A&A...549A.137I}; \citeads{2015ApJ...804...49B}). Because the star is too far away to have a useful {\it Gaia} DR2 parallax ($-0.0117 \pm 0.0582$\,mas), kinematic analyses still have to rely on alternative distance estimates. By making use of recent photometric measurements, we revisit the spectrophotometric distance of HVS\,3. Combined with proper motions from {\it Gaia} DR2, its place of origin can finally be constrained.   
\subsection{Spectrophotometric distance of HVS 3 revisited}
\begin{figure*}
\begin{center}
\sidecaption
\includegraphics[width=12.1cm]{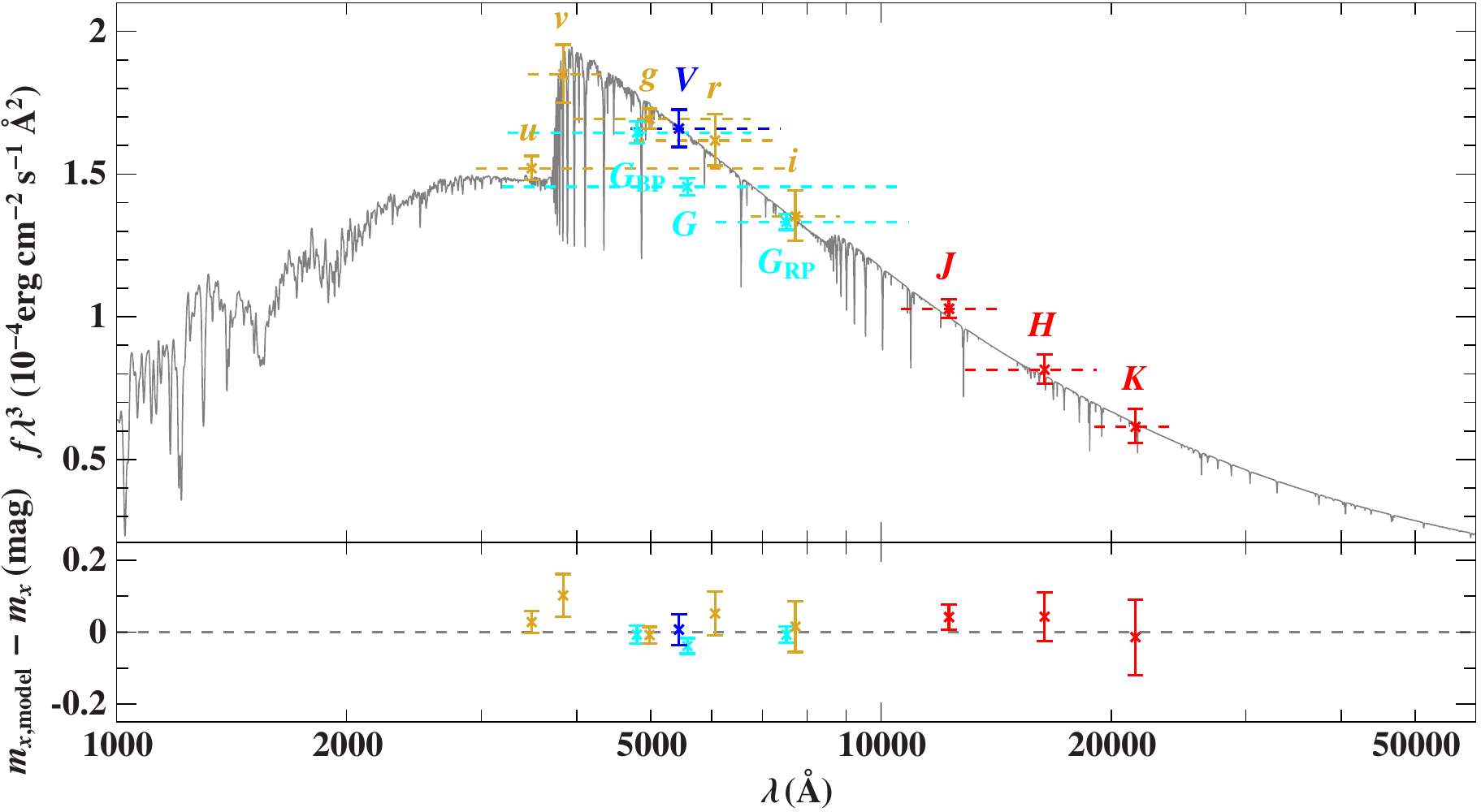}
\caption{Comparison of synthetic and observed photometry for HVS\,3: The \textit{top panel} shows the SED. The colored data points (APASS DR9: blue, SkyMapper DR1: yellow, {\it Gaia} DR2: cyan, VHS DR5: red) are filter-averaged fluxes that were converted from observed magnitudes (the respective filter widths are indicated by the dashed horizontal lines), while the gray solid line represents a model flux distribution that is based on the parameters ($T_\mathrm{eff}=23\,000$\,K, $\log(g)=3.95$) derived by \citetads{2008A&A...480L..37P}. The flux is multiplied with the wavelength to the power of three to reduce the steep slope of the SED on such a wide wavelength range. The residual panel at the \textit{bottom} shows the differences between synthetic and observed magnitudes. Note that only the angular diameter and the color excess were fitted.
\label{fig:sed_fit}}
\end{center}
\end{figure*}
Photometric observations of HVS\,3 were scarce. \citetads{2005ApJ...634L.181E} and \citetads{2008A&A...480L..37P} had to rely on a single photographic magnitude ($V=16.2 \pm 0.2$\,mag) from the Hamburg ESO Survey and to assume zero interstellar reddening to derive the spectrophotometric distance of $61$\,kpc. Only later did photoelectric measurements become available (\citeads{2008ApJ...675L..77B}; \citeads{2013MNRAS.431..240O}; \citeads{2015MNRAS.453.1879K}; APASS: \citeads{2016yCat.2336....0H}). SkyMapper \citepads{2018PASA...35...10W}, {\it Gaia} \citepads{2018A&A...616A...4E}, and the VISTA Hemisphere Survey (VHS, \citeads{2013Msngr.154...35M}) now provide high-precision optical and infrared photometry to construct the SED, which we used to determine the stellar angular diameter $\Theta$ and interstellar reddening in form of the color excess $E(B-V)$ (for details, see \citetalias{2018A&A...615L...5I} and \citeads{2018OAst...27...35H}). To this end, we computed a synthetic SED based on the atmospheric parameters derived by \citetads{2008A&A...480L..37P} from high-resolution spectroscopy and fit it to the photometric observations by varying $\Theta$ and $E(B-V)$. The resulting synthetic SED matches optical and infrared photometry very well (see Fig.~\ref{fig:sed_fit}) and yields a color excess $E(B-V) = 0.007^{+0.009}_{-0.007}$\,mag that is consistent with zero. Assuming a mass of $M=9.1\,M_\odot$ \citepads{2008A&A...480L..37P}, we can compute the distance $d$ by expressing the stellar radius $R$ in terms of the surface gravity $g = MG/R^2$ and inserting it in the defintion for the angular diameter $\theta = 2R/d$. Our revised value of $d=62.3 \pm 7.7$\,kpc is consistent with the estimate of $61\pm9$\,kpc \citepads{2008A&A...480L..37P} and the inconclusive {\it Gaia} parallax. 
\subsection{Past trajectories of HVS 3 and LMC}
\begin{figure*}
\includegraphics[width=1\textwidth]{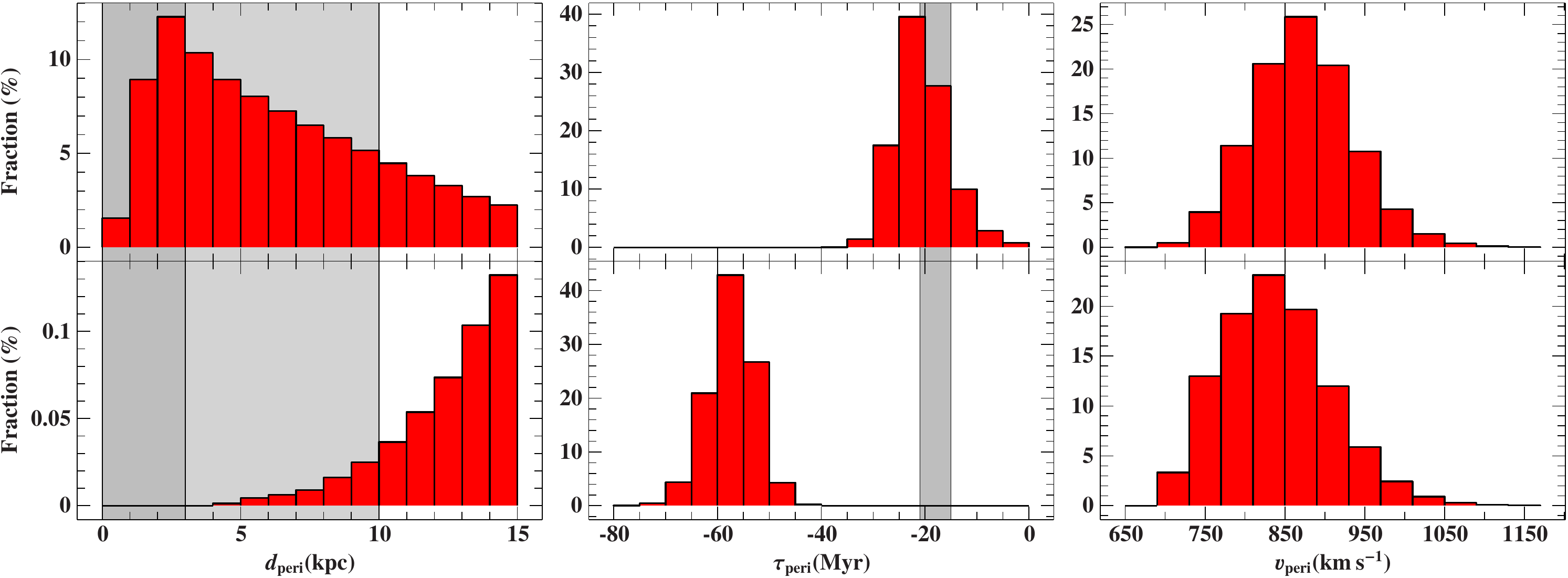}
\caption{Histograms showing the distribution of distances $d_{\mathrm{peri}}$, times $\tau_{\mathrm{peri}}$, and relative velocities $\varv_{\mathrm{peri}}$ at periastron of HVS\,3 with respect to the LMC (\textit{upper panel}) and the GC (\textit{lower panel}) for Model~I of \citetads{2013A&A...549A.137I}. The gray shaded areas mark regions with $d_{\mathrm{peri}} \le R^{\mathrm{LMC}}_{\mathrm{cen}} = 3\,\mathrm{kpc}$, $d_{\mathrm{peri}} \le R^{\mathrm{LMC}}_{\mathrm{out}} = 10\,\mathrm{kpc}$, and $15\,\mathrm{Myr} \le |\tau_{\mathrm{peri}}| \le 21\,\mathrm{Myr}$. The last is the lifetime of HVS\,3 assuming a single-star nature \citepads{2008A&A...480L..37P}.}
\label{fig:histograms}
\end{figure*}
To test an LMC origin of HVS\,3, we computed $10^6$ orbit pairs in a Monte Carlo simulation. The input parameters of HVS\,3 are listed in Table~\ref{table:kinematic_short}. The distance employed for the LMC is $50.1 \pm 2.4$\,kpc \citepads{2001ApJ...553...47F}, the radial velocity is $262.2 \pm 3.4$\,km\,s$^{-1}$ \citepads{2002AJ....124.2639V}, and proper motion components are $\mu_{\alpha}\cos\delta = 1.910 \pm 0.020\,\mathrm{mas} \,\mathrm{yr}^{-1}$ and $\mu_{\delta} = 0.229 \pm 0.047\,\mathrm{mas} \,\mathrm{yr}^{-1}$ \citepads{2013ApJ...764..161K}, which are in excellent agreement with {\it Gaia} measurements \citepads{2018A&A...616A..12G}. To model the gravitational attraction of the LMC on HVS\,3, we followed our approach in \citetads{2013A&A...549A.137I} and used a moving Plummer potential with a mass of $2 \times 10^{10}\,M_{\odot}$ and a radius of 3\,kpc. The resulting distributions of distance $d_{\mathrm{peri}}$, time $\tau_{\mathrm{peri}}$, and relative velocity $\varv_{\mathrm{peri}}$ at periastron are relatively insensitive to the adopted masses of the Milky Way and of the LMC and are therefore only shown for Model~I in Fig.~\ref{fig:histograms}. For comparison, we also show the same quantities with respect to the GC. The results clearly demonstrate that HVS\,3 and the LMC had a very close encounter just at the time when HVS\,3 was born ($d_\mathrm{peri} = 6^{+7}_{-4}$\,kpc, $\tau_\mathrm{peri} = 21^{+5}_{-6}$\,Myr). However, the ejection velocity of $\varv_\mathrm{peri} = 870\pm70$\,km\,s$^{-1}$ is higher than for any other star in our sample. These results are in agreement with those of \citetads{2018arXiv180410197E}, who used different Galactic potentials (\citeads{2015ApJS..216...29B}; \citeads{2017MNRAS.465...76M}) and LMC parameters.    
\section{Summary and discussion}\label{sect:discussion}
\begin{figure*}
\begin{center}
\includegraphics[width=0.99\textwidth]{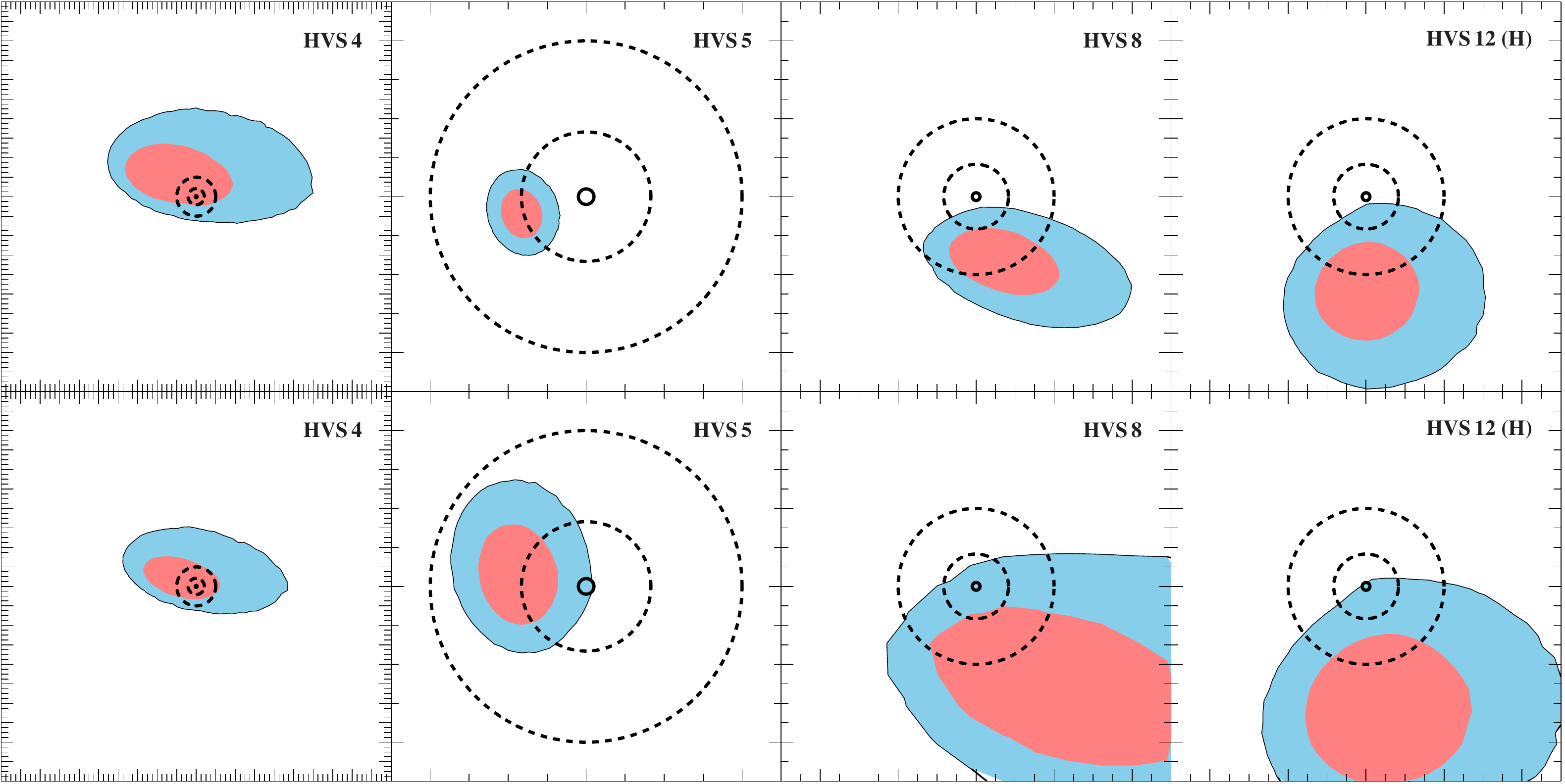}
\caption{Effect of different distance estimates on the Galactic plane-crossing locations for four diverging cases (see Fig.~\ref{fig:dist_comp}). The \textit{upper panel} is based on spectroscopic distances from \citetalias{2018A&A...615L...5I}, that is, it shows exactly the same as Fig.~\ref{fig:grid}, while spectroscopic distances from \citetads{2015ApJ...804...49B} are used in the \textit{lower panel} (all other parameters are the same).}
\label{fig:distance_differences}
\end{center}
\end{figure*}
We selected 15 high-velocity stars that have the best pre-{\it Gaia} proper motions, measured with the HST. The second {\it Gaia} data release showed that for 4 of them (HVS\,1, HVS\,10, HVS\,12, and HVS\,13), the HST measurements remain the best. For 3 stars (HVS\,4, HVS\,6, and HVS\,9), data from HST and {\it Gaia} DR2 are more or less consistent and at a similar level of precision. Most important are the {\it Gaia} DR2 proper motions for HVS\,3, HVS\,5, HVS\,7, HVS\,8, B434, B485, and B733 because they are superior to, while still consistent with, the HST measurements. For B711, the statistical uncertainties of the {\it Gaia} DR2 proper motions are considerably smaller than those from the HST, but both are seriously different, which hints at an as yet unidentified systematic uncertainty for that star. Because HVSs are widely assumed to originate in the GC through the Hills mechanism, we focused on identifying the place of origin in the Galactic plane. By modeling the trajectories of the stars in three different Milky Way mass models, we derived their Galactic plane-crossing properties. 

\citetads{2018arXiv180504184B} carried out similar kinematic calculations based on the same {\it Gaia} DR2 proper motions and a Galactic potential \citepads{2014ApJ...793..122K} whose total mass is close to that of Model~II employed in this study. Therefore, similar results are in principle expected. However, there are two main differences: First, \citetads{2018arXiv180504184B} relied on distances from \citetads{2015ApJ...804...49B}, while we used our revised distances from \citetalias{2018A&A...615L...5I}. As shown in Section~\ref{subsection:spectrophotometric_distances}, the two distance estimates differ markedly for five stars. Our smaller distances for three of them (HVS\,5, HVS\,8, and HVS\,10) allow their spatial origin to be constrained with much higher precision, enabling us to rule out an ejection from the GC at more than $2\sigma$ confidence (see Fig.~\ref{fig:distance_differences}). Second, and more importantly, the outcome of the kinematic calculations is viewed from quite different perspectives. While \citetads{2018arXiv180504184B} computed probabilities for a Galactic disk origin by assuming that the most likely place of origin is that with the lowest ejection velocity, our approach is a straightforward interpretation of the plane-crossing properties, which then hints at the existence of an as yet unknown or neglected but powerful disk-ejection mechanism. In the following, we summarize and discuss our findings.

For 6 stars (HVS\,1, HVS\,4, HVS\,6, HVS\,9, HVS\,10, and HVS\,13), the available astrometry is (still) insufficient to place meaningful constraints on their place of origin within the Galactic plane. Hence, a GC origin remains a valid option for them, in particular for HVS\,1 and HVS\,4, whose ejection velocities ($750 \pm 90$\,km\,s$^{-1}$, $840^{+\phantom{0}70}_{-130}$\,km\,s$^{-1}$) are so high that the Hills mechanism is the most likely scenario. For 5 unbound (HVS\,3, HVS\,5, HVS\,7, HVS\,8, and HVS\,12) and 4 bound (B434, B485, B711, and B733) stars, however, astrometric data were sufficiently precise to rule out the GC as the place of origin. Consequently, the Galactic plane-crossing properties of our sample suggest that a large portion of the high-velocity stars are actually runaways launched at very high velocities from the Galactic disk rather than from the GC. Until recently, only 2 unbound runaway B-stars, so-called hyper-runaway stars, were known (\citeads{2008A&A...483L..21H}; \citeads{2010ApJ...711..138I}). Three new members (HVS\,5, HVS\,7, and HVS\,8) from this work and one (LAMOST-HVS\,4) from \citetads{2018AJ....156...87L} are now added to form a group of six. Studying a sample of 96 MS B-type runaway stars, \citetads{2011MNRAS.411.2596S} discovered that the distribution of their ejection velocities $\varv_{\rm{ej}}$ is bimodal. A small group of 11 ``high velocity'' stars ($\varv_{\rm{ej}}=400$--$500$\,km\,s$^{-1}$) is well separated from the ``low velocity'' population with $\varv_{\rm{ej}}<300$\,km\,s$^{-1}$. Taking into account the runaway stars B434, B485, and B733, the number of ``high velocity'' runaway B-stars increases to 14. 

Two mechanisms have been proposed decades ago for ejecting runaway stars. \citetads{1961BAN....15..265B} proposed a binary supernova ejection (BSE) scenario, where the secondary star of a close binary system is ejected when the binary system is disrupted by the core-collapse of the more massive primary. Population synthesis models suggest that only for a small fraction, that is, less than 1\%, ejection velocities above 200\,km\,s$^{-1}$ \citepads{2000ApJ...544..437P} or even above 60\,km\,s$^{-1}$ \citepads{2018arXiv180409164R} can be reached. Hence, hardly any are expected in excess of 400\,km\,s$^{-1}$, as found in this study. However, in extremely rare cases when a very close binary is disrupted, higher velocities can be achieved. \citetads{2015MNRAS.448L...6T} studied the ejection of late B-type stars of 3.5\,$M_\odot$, which is similar to the objects considered here, and found an upper BSE limit of 540\,km\,s$^{-1}$ in the most favorable conditions. Such stars could reach their local Galactic escape velocity if the ejection happens to occur in the direction of Galactic rotation. Interestingly, we found three (HVS\,5, HVS\,7, HVS\,12) out of five unbound and two (B434, B711) out of four bound stars to be disk runaways with ejection velocities close to or even above (HVS\,5 and B434) that limit. Therefore, BSE cannot satisfactorily explain most of our extreme disk runaway stars.

The second mechanism to create runaway stars is dynamical ejection (DE) via three- or four-body interactions in a dense stellar environment \citepads{1967BOTT....4...86P}. N-body simulations of the dynamical ejection of massive stars from moderately massive star clusters by  \citetads{2016A&A...590A.107O} show that only very few stars with masses lower than 5\,$M_\odot$ are ejected from a young star cluster with 3000\,$M_\odot$ at more than 100\,km\,s$^{-1}$ (see their Fig.~4). \citetads{2012ApJ...751..133P} carried out similar N-body simulations for star cluster dynamics and concluded that ``the ejection rate of hyper-runaways, with velocities $>$300\,km\,s$^{-1}$, appears to be too low to explain a significant fraction of the observed HVSs in the Galactic halo and could at most explain a small fraction of the observed bound HVSs''. Thus, the ``classical'' dynamical ejection from a star cluster does not seem to be a viable scenario either.

A combination of the DE and BSE mechanisms is also conceivable. A massive, tightly bound binary may be ejected from a dense, massive stellar cluster, and thereafter, the primary explodes as a core-collapse supernova. Numerical experiments by \citetads{2012ApJ...751..133P} indicate that no binaries are ejected by DE with velocities in excess of 150\,km\,s$^{-1}$. Hence, the combined DE/BSE scenario would shift the BSE speed limit by such an amount in the most favorable, though very unlikely, case of a perfect alignment between ejection vectors. This scenario could explain the hyper-runaway star HD\,271791 \citepads{2008A&A...483L..21H}, which shows atmospheric chemical peculiarities indicative of supernova debris accretion \citepads{2008ApJ...684L.103P}.

A more realistic explanation for our extreme disk-runaway objects, however, is offered by dynamical interactions involving massive perturbers. Three possible channels for extreme dynamical ejection events were discussed by \citetads{2009MNRAS.395L..85G}: (i)~The break-up of unstable, close triple systems, (ii)~the interaction of two massive close binaries, and (iii)~close encounters between massive close binaries with a very massive star. (i)~For a triple system composed of an inner binary with two MS stars of 50\,$M_\odot$ with an orbital separation as small as 40\,$R_\odot$, which is close to filling their Roche lobes, and an outer component of 10\,$M_\odot$, \citetads{2009MNRAS.395L..85G} estimated that the latter may be ejected at a velocity as high as $\approx$ 800\,km\,s$^{-1}$. (ii)~Although runaways from binary-binary ejections are likely to be ejected at velocities similar to the stars' orbital velocities \citepads{1990AJ.....99..608L}, the interaction of massive close binaries could eject stars at velocities as high as the escape velocity from the surface of the most massive component in the binaries \citepads{1991AJ....101..562L}, which could exceed 1000\,km\,s$^{-1}$ if stars of 20--40\,$M_\odot$ were involved \citepads{2009MNRAS.395L..85G}. (iii) Runaways from three-body interactions between a massive close binary and a very massive star or an intermediate-mass black hole of 1000--10\,000\,$M_\odot$ were studied by \citeauthor{2008MNRAS.385..929G} (\citeyearads{2008MNRAS.385..929G}, \citeyearads{2009MNRAS.396..570G}). In both cases, high stellar densities of $10^6$ to $10^7$ stars per pc$^{3}$ are required, which may occur when a young massive cluster forms in a core-collapse phase, but have never been observed so far. If the high-velocity stars were indeed ejected during cluster formation, their stellar ages would be expected to be almost identical to their times of flight. However, this it not the case for the extreme disk-runaway stars in our sample. Except for HVS\,12, the stellar lifetimes are significantly longer than the flight times (see Table~\ref{table:kinematic_short} and Fig.~\ref{fig:tof_age}). {\it Gaia} will improve the census of Milky Way open clusters \citepads{2018arXiv180508726C} and may finally allow the parent clusters of these stars to be identified. 

For the most massive HVS, HVS\,3, we revised the spectrophotometric distance ($62.3\pm7.7$\,kpc) using optical and infrared photometry and excluded a place of origin in the Galactic disk. Previous indications that the star originates in the LMC are confirmed here. The most likely place of origin lies only $6^{+7}_{-4}$\,kpc from the center of the LMC. Its ejection velocity is as high as $870\pm70$\,km\,s$^{-1}$, well above the classical speed limits for BSE and DE. \citetads{2007MNRAS.376L..29G} suggested that HVS\,3 was dynamically ejected from a young massive star cluster in the LMC, but \citetads{2018arXiv180410197E} considered the very high ejection velocity as strong evidence for the ejection via the Hills mechanism, and therefore as evidence for the presence of an as yet undiscovered massive black hole in the LMC.
\begin{acknowledgements}
We thank John E.\ Davis for the development of the {\sc slxfig} module we used to prepare some figures in this paper. 
This work has made use of data from the European Space Agency (ESA) mission
{\it Gaia} (\url{https://www.cosmos.esa.int/gaia}), processed by the {\it Gaia}
Data Processing and Analysis Consortium (DPAC,
\url{https://www.cosmos.esa.int/web/gaia/dpac/consortium}). Funding for the DPAC
has been provided by national institutions, in particular the institutions
participating in the {\it Gaia} Multilateral Agreement. 
The national facility capability for SkyMapper has been funded through ARC LIEF grant LE130100104 from the Australian Research Council, awarded to the University of Sydney, the Australian National University, Swinburne University of Technology, the University of Queensland, the University of Western Australia, the University of Melbourne, Curtin University of Technology, Monash University and the Australian Astronomical Observatory. SkyMapper is owned and operated by The Australian National University's Research School of Astronomy and Astrophysics. The survey data were processed and provided by the SkyMapper Team at ANU. The SkyMapper node of the All-Sky Virtual Observatory (ASVO) is hosted at the National Computational Infrastructure (NCI). Development and support the SkyMapper node of the ASVO has been funded in part by Astronomy Australia Limited (AAL) and the Australian Government through the Commonwealth's Education Investment Fund (EIF) and National Collaborative Research Infrastructure Strategy (NCRIS), particularly the National eResearch Collaboration Tools and Resources (NeCTAR) and the Australian National Data Service Projects (ANDS). 
Based on observations obtained as part of the VISTA Hemisphere Survey, ESO Program, 179.A-2010 (PI: McMahon). 
\end{acknowledgements}
\bibliographystyle{aa}

%
\onecolumn
\begin{appendix}
\section{Additional tables and figures}
\begin{table*}[h]
\begin{center}
\footnotesize
\setlength{\tabcolsep}{0.09cm}
\renewcommand{\arraystretch}{1.3}
\caption{\label{table:kinematic_parameters_I} Kinematic parameters of the program stars for Model~I of \citetads{2013A&A...549A.137I}.}
\begin{tabular}{lrrrrrrrrrrrrrrrrrrrrr}
\hline\hline
 & $x$ & $y$ & $z$ & & $\varv_x$ & $\varv_y$ & $\varv_z$ & $\varv_{\mathrm{Grf}}$ & $\varv_{\mathrm{Grf}}-\varv_{\mathrm{esc}}$ & $P_{\mathrm{b}}$ & $x_{\mathrm{p}}$ & $y_{\mathrm{p}}$ & $z_{\mathrm{p}}$ & $r_{\mathrm{p}}$ & & $\varv_{x\mathrm{,p}}$ & $\varv_{y\mathrm{,p}}$ & $\varv_{z\mathrm{,p}}$ & $\varv_{\mathrm{Grf,p}}$ &  $\varv_{\mathrm{ej,p}}$ & $\tau_{\mathrm{flight,p}}$ \\
\cline{2-4} \cline{6-10} \cline{12-15} \cline{17-21}
& \multicolumn{3}{c}{(kpc)} & & \multicolumn{5}{c}{$(\mathrm{km\,s^{-1}})$} & (\%) & \multicolumn{4}{c}{(kpc)} & & \multicolumn{5}{c}{$(\mathrm{km\,s^{-1}})$} & (Myr) \\
\hline \hline
 HVS\,1 (H) & $-65.9$ & $-62.4$ &  $51.6$ &   & $-420$ & $-310$ &  $440$ &  $690$ &  $330$ &   $0$ & $-16.0$ & $-24.3$ &  $0.0$ &  $34.2$ &   & $-470$ & $-380$ &  $470$ &  $770$ & $750$ & $112$ \\ Stat. & $^{+5.4}_{-8.9}$ &             $^{+5.8}_{-9.6}$ &             $^{+8.0}_{-4.8}$ &   &           $^{+110}_{-100}$ &   $^{+80}_{-90}$ & $^{+110}_{-110}$ &             $^{+40}_{-20}$ &             $^{+40}_{-30}$ & \ldots &             $^{+22.7}_{-17.5}$ &           $^{+18.3}_{-14.1}$ & $^{+0.1}_{-0.1}$ &           $^{+16.8}_{-16.1}$ &   & $^{+110}_{-\phantom{0}80}$ & $^{+100}_{-\phantom{0}90}$ & $^{+100}_{-\phantom{0}60}$ &             $^{+30}_{-20}$ &             $^{+90}_{-90}$ &             $^{+34}_{-22}$ \\
     HVS\,3 & $-14.1$ & $-46.7$ & $-40.8$ &   & $-670$ & $-310$ & $-360$ &  $820$ &  $400$ &   $0$ &  $55.7$ &  $-9.9$ &  $0.0$ &  $56.8$ &   & $-620$ & $-370$ & $-400$ &  $820$ & $720$ & $104$ \\ Stat. & $^{+0.8}_{-0.8}$ &             $^{+5.8}_{-5.9}$ &             $^{+5.1}_{-5.1}$ &   & $^{+100}_{-\phantom{0}90}$ &   $^{+30}_{-30}$ &   $^{+40}_{-30}$ &             $^{+70}_{-70}$ &             $^{+80}_{-80}$ & \ldots &             $^{+22.2}_{-17.5}$ &             $^{+4.6}_{-4.2}$ & $^{+0.1}_{-0.1}$ &           $^{+21.8}_{-17.1}$ &   & $^{+\phantom{0}90}_{-100}$ &             $^{+30}_{-30}$ &             $^{+30}_{-30}$ &             $^{+70}_{-50}$ &             $^{+90}_{-70}$ &             $^{+20}_{-17}$ \\
     HVS\,4 & $-64.2$ & $-14.7$ &  $52.9$ &   & $-360$ & $-310$ &  $360$ &  $630$ &  $240$ &   $0$ & $-12.9$ &  $25.9$ &  $0.0$ &  $45.8$ &   & $-430$ & $-280$ &  $440$ &  $680$ & $840$ & $129$ \\ Stat. & $^{+5.2}_{-6.1}$ &             $^{+1.4}_{-1.6}$ &             $^{+5.9}_{-4.8}$ &   &           $^{+190}_{-180}$ & $^{+180}_{-200}$ & $^{+160}_{-170}$ & $^{+120}_{-\phantom{0}60}$ & $^{+120}_{-\phantom{0}60}$ & \ldots &             $^{+46.9}_{-32.5}$ &           $^{+22.9}_{-18.5}$ & $^{+0.1}_{-0.1}$ &           $^{+27.1}_{-22.6}$ &   &           $^{+210}_{-100}$ &           $^{+160}_{-200}$ &           $^{+120}_{-100}$ &             $^{+90}_{-30}$ & $^{+\phantom{0}70}_{-130}$ &             $^{+56}_{-35}$ \\
     HVS\,5 & $-28.6$ &  $13.5$ &  $19.5$ &   & $-400$ &  $330$ &  $400$ &  $650$ &  $180$ &   $0$ &  $-8.5$ &  $-2.4$ &  $0.0$ &   $9.1$ &   & $-500$ &  $340$ &  $450$ &  $760$ & $640$ &  $46$ \\ Stat. & $^{+1.6}_{-2.1}$ &             $^{+1.5}_{-1.1}$ &             $^{+2.0}_{-1.6}$ &   &             $^{+30}_{-30}$ &   $^{+40}_{-40}$ &   $^{+30}_{-30}$ &             $^{+10}_{-10}$ &             $^{+10}_{-10}$ & \ldots &               $^{+1.8}_{-1.8}$ &             $^{+2.2}_{-2.1}$ & $^{+0.1}_{-0.1}$ &             $^{+1.7}_{-1.8}$ &   &             $^{+30}_{-40}$ &             $^{+40}_{-30}$ &             $^{+20}_{-20}$ &             $^{+20}_{-20}$ &             $^{+50}_{-40}$ &               $^{+4}_{-5}$ \\
     HVS\,6 & $-21.6$ & $-26.1$ &  $49.7$ &   & $-160$ & $-170$ &  $460$ &  $550$ &  $120$ &   $0$ &  $-3.8$ &  $-6.1$ &  $0.0$ &  $19.3$ &   & $-220$ & $-250$ &  $570$ &  $650$ & $680$ &  $96$ \\ Stat. & $^{+1.7}_{-1.6}$ &             $^{+3.3}_{-3.0}$ &             $^{+5.8}_{-6.2}$ &   &           $^{+170}_{-170}$ & $^{+130}_{-140}$ &   $^{+90}_{-90}$ &             $^{+60}_{-30}$ &             $^{+70}_{-30}$ & \ldots &             $^{+17.9}_{-16.2}$ &           $^{+15.5}_{-13.1}$ & $^{+0.1}_{-0.1}$ &           $^{+13.9}_{-10.2}$ &   & $^{+160}_{-\phantom{0}90}$ & $^{+130}_{-\phantom{0}70}$ &             $^{+50}_{-60}$ &             $^{+50}_{-30}$ &             $^{+90}_{-80}$ &             $^{+21}_{-17}$ \\
     HVS\,7 & $-11.1$ & $-25.4$ &  $40.8$ &   & $-200$ &   $-0$ &  $450$ &  $500$ &   $50$ &   $5$ &   $6.3$ & $-22.1$ &  $0.0$ &  $24.1$ &   & $-200$ & $-100$ &  $510$ &  $570$ & $530$ &  $82$ \\ Stat. & $^{+0.2}_{-0.3}$ &             $^{+2.0}_{-2.4}$ &             $^{+3.7}_{-3.2}$ &   & $^{+\phantom{0}90}_{-100}$ &   $^{+50}_{-50}$ &   $^{+40}_{-30}$ &             $^{+50}_{-40}$ &             $^{+50}_{-40}$ & \ldots &               $^{+8.2}_{-7.2}$ &             $^{+4.8}_{-4.9}$ & $^{+0.1}_{-0.1}$ &             $^{+5.4}_{-4.8}$ &   &             $^{+70}_{-80}$ &             $^{+70}_{-60}$ &             $^{+30}_{-20}$ &             $^{+20}_{-30}$ &             $^{+30}_{-30}$ &   $^{+10}_{-\phantom{0}8}$ \\
     HVS\,8 & $-30.3$ & $-13.5$ &  $26.9$ &   & $-420$ &   $80$ &  $260$ &  $500$ &   $40$ &  $16$ &   $9.3$ & $-16.4$ &  $0.0$ &  $19.8$ &   & $-450$ &  $-40$ &  $350$ &  $570$ & $450$ &  $87$ \\ Stat. & $^{+2.2}_{-2.6}$ &             $^{+1.4}_{-1.7}$ &             $^{+3.2}_{-2.6}$ &   &             $^{+70}_{-60}$ &   $^{+60}_{-70}$ &   $^{+50}_{-50}$ &             $^{+50}_{-40}$ &             $^{+60}_{-50}$ & \ldots &   $^{+11.3}_{-\phantom{0}8.0}$ &             $^{+5.2}_{-6.1}$ & $^{+0.1}_{-0.1}$ &             $^{+9.8}_{-6.3}$ &   &             $^{+40}_{-40}$ &             $^{+70}_{-80}$ &             $^{+30}_{-40}$ &             $^{+20}_{-10}$ &             $^{+40}_{-30}$ &             $^{+18}_{-14}$ \\
     HVS\,9 & $-28.8$ & $-43.0$ &  $46.5$ &   &  $-40$ & $-170$ &  $480$ &  $570$ &  $160$ &   $0$ & $-22.6$ & $-24.1$ &  $0.0$ &  $41.8$ &   & $-100$ & $-230$ &  $520$ &  $620$ & $690$ &  $90$ \\ Stat. & $^{+2.2}_{-1.9}$ &             $^{+4.6}_{-4.0}$ &             $^{+4.4}_{-4.9}$ &   &           $^{+250}_{-250}$ & $^{+170}_{-170}$ & $^{+150}_{-150}$ & $^{+140}_{-\phantom{0}80}$ & $^{+140}_{-\phantom{0}90}$ & \ldots &             $^{+25.0}_{-22.2}$ &           $^{+24.5}_{-17.0}$ & $^{+0.1}_{-0.1}$ &           $^{+16.3}_{-16.9}$ &   &           $^{+260}_{-200}$ &           $^{+180}_{-160}$ &           $^{+130}_{-100}$ & $^{+110}_{-\phantom{0}50}$ &           $^{+110}_{-120}$ &             $^{+34}_{-21}$ \\
HVS\,10 (H) & $-13.0$ & $-12.6$ &  $52.5$ &   & $-200$ & $-110$ &  $380$ &  $450$ &   $20$ &  $29$ &   $9.5$ &   $1.2$ &  $0.0$ &  $14.9$ &   & $-160$ & $-130$ &  $540$ &  $580$ & $600$ & $114$ \\ Stat. & $^{+0.5}_{-0.6}$ &             $^{+1.3}_{-1.4}$ &             $^{+6.1}_{-5.3}$ &   &           $^{+110}_{-100}$ & $^{+110}_{-110}$ &   $^{+30}_{-30}$ &             $^{+60}_{-30}$ &             $^{+60}_{-40}$ & \ldots &   $^{+12.7}_{-\phantom{0}9.5}$ &           $^{+12.5}_{-10.3}$ & $^{+0.1}_{-0.1}$ & $^{+12.8}_{-\phantom{0}8.3}$ &   &             $^{+50}_{-70}$ &             $^{+50}_{-50}$ &             $^{+70}_{-50}$ &             $^{+60}_{-20}$ &             $^{+70}_{-40}$ &             $^{+18}_{-14}$ \\
HVS\,12 (H) & $-20.6$ & $-29.0$ &  $41.0$ &   & $-240$ &  $-20$ &  $430$ &  $500$ &   $60$ &   $8$ &   $1.7$ & $-23.8$ &  $0.0$ &  $25.6$ &   & $-260$ & $-110$ &  $480$ &  $570$ & $510$ &  $88$ \\ Stat. & $^{+1.4}_{-2.2}$ &             $^{+3.4}_{-5.1}$ &             $^{+7.2}_{-4.9}$ &   &             $^{+90}_{-90}$ &   $^{+80}_{-70}$ &   $^{+50}_{-60}$ &             $^{+60}_{-50}$ &             $^{+60}_{-50}$ & \ldots &   $^{+10.4}_{-\phantom{0}7.9}$ &             $^{+8.0}_{-8.7}$ & $^{+0.1}_{-0.1}$ &             $^{+9.1}_{-7.7}$ &   &             $^{+70}_{-70}$ & $^{+\phantom{0}90}_{-100}$ &             $^{+50}_{-30}$ &             $^{+30}_{-20}$ &             $^{+40}_{-30}$ &             $^{+19}_{-14}$ \\
HVS\,13 (H) & $-27.4$ & $-57.3$ &  $73.6$ &   & $-550$ &  $-50$ &  $360$ &  $690$ &  $310$ &   $0$ &  $68.9$ & $-42.0$ &  $0.0$ &  $89.3$ &   & $-500$ & $-110$ &  $430$ &  $680$ & $610$ & $179$ \\ Stat. & $^{+2.6}_{-3.6}$ & $^{+\phantom{0}7.7}_{-10.8}$ & $^{+13.8}_{-\phantom{0}9.9}$ &   &           $^{+190}_{-200}$ & $^{+170}_{-160}$ & $^{+130}_{-130}$ &           $^{+170}_{-150}$ &           $^{+190}_{-150}$ & \ldots &             $^{+64.4}_{-38.5}$ &           $^{+37.5}_{-31.2}$ & $^{+0.1}_{-0.1}$ &           $^{+57.6}_{-35.9}$ &   &           $^{+180}_{-200}$ &           $^{+170}_{-150}$ &           $^{+100}_{-100}$ &           $^{+160}_{-100}$ &           $^{+170}_{-100}$ &             $^{+72}_{-44}$ \\
       B434 & $-16.0$ & $-22.4$ &  $32.9$ &   &  $120$ & $-280$ &  $220$ &  $380$ &  $-80$ &  $92$ & $-24.2$ &  $13.2$ &  $0.0$ &  $29.0$ &   &  $-10$ & $-280$ &  $320$ &  $430$ & $590$ & $118$ \\ Stat. & $^{+0.7}_{-0.9}$ &             $^{+2.1}_{-2.6}$ &             $^{+3.9}_{-3.0}$ &   &             $^{+80}_{-70}$ &   $^{+60}_{-60}$ &   $^{+40}_{-50}$ &             $^{+50}_{-40}$ &             $^{+60}_{-40}$ & \ldots &   $^{+\phantom{0}9.0}_{-10.7}$ & $^{+13.1}_{-\phantom{0}8.7}$ & $^{+0.1}_{-0.1}$ & $^{+13.2}_{-\phantom{0}9.3}$ &   & $^{+100}_{-\phantom{0}90}$ &             $^{+30}_{-40}$ &             $^{+30}_{-40}$ &             $^{+20}_{-10}$ &             $^{+20}_{-20}$ &             $^{+26}_{-19}$ \\
       B485 & $-26.7$ &  $-5.9$ &  $27.2$ &   & $-330$ &  $140$ &  $270$ &  $450$ &  $-20$ &  $89$ &   $4.4$ & $-14.8$ &  $0.0$ &  $15.6$ &   & $-390$ &   $20$ &  $380$ &  $540$ & $420$ &  $83$ \\ Stat. & $^{+1.0}_{-2.1}$ &             $^{+0.3}_{-0.7}$ &             $^{+3.0}_{-1.5}$ &   &             $^{+30}_{-30}$ &   $^{+20}_{-20}$ &   $^{+20}_{-20}$ &             $^{+20}_{-20}$ &             $^{+20}_{-20}$ & \ldots &               $^{+3.7}_{-2.7}$ &             $^{+1.6}_{-2.1}$ & $^{+0.1}_{-0.1}$ &             $^{+2.9}_{-2.1}$ &   &             $^{+10}_{-10}$ &             $^{+20}_{-30}$ &             $^{+10}_{-20}$ &             $^{+10}_{-10}$ &             $^{+20}_{-10}$ &   $^{+11}_{-\phantom{0}6}$ \\
       B711 &   $3.7$ &   $0.5$ &  $25.8$ &   &  $390$ &   $50$ &  $130$ &  $420$ &  $-90$ &  $99$ & $-36.0$ &  $-4.5$ &  $0.0$ &  $36.4$ &   &  $270$ &   $30$ &  $270$ &  $380$ & $440$ & $113$ \\ Stat. & $^{+1.4}_{-0.9}$ &             $^{+0.1}_{-0.1}$ &             $^{+2.8}_{-2.0}$ &   &             $^{+40}_{-40}$ &   $^{+30}_{-40}$ &   $^{+20}_{-20}$ &             $^{+30}_{-30}$ &             $^{+40}_{-30}$ & \ldots &   $^{+\phantom{0}7.1}_{-10.6}$ &             $^{+3.3}_{-2.7}$ & $^{+0.1}_{-0.1}$ & $^{+10.5}_{-\phantom{0}7.0}$ &   &             $^{+40}_{-40}$ &             $^{+30}_{-20}$ &             $^{+20}_{-30}$ &             $^{+20}_{-10}$ &             $^{+10}_{-10}$ &             $^{+22}_{-14}$ \\
   B711 (H) &   $3.7$ &   $0.5$ &  $25.8$ &   &  $-90$ &  $340$ &  $350$ &  $510$ &    $0$ &  $48$ &   $7.8$ & $-18.8$ &  $0.0$ &  $21.8$ &   &  $-30$ &  $250$ &  $460$ &  $530$ & $600$ &  $61$ \\ Stat. & $^{+1.4}_{-1.0}$ &             $^{+0.1}_{-0.1}$ &             $^{+2.8}_{-2.1}$ &   &           $^{+130}_{-120}$ & $^{+130}_{-140}$ &   $^{+60}_{-60}$ &           $^{+120}_{-110}$ &           $^{+130}_{-100}$ & \ldots &               $^{+7.2}_{-7.2}$ &             $^{+8.3}_{-9.8}$ & $^{+0.1}_{-0.1}$ &             $^{+9.3}_{-8.0}$ &   & $^{+\phantom{0}90}_{-120}$ &           $^{+150}_{-140}$ &             $^{+50}_{-60}$ & $^{+100}_{-\phantom{0}50}$ &             $^{+90}_{-50}$ &   $^{+11}_{-\phantom{0}8}$ \\
       B733 &  $-5.5$ &   $3.3$ &   $8.9$ &   &  $230$ &  $190$ &  $350$ &  $460$ & $-120$ & $100$ &  $-9.9$ &  $-1.3$ &  $0.0$ &  $10.0$ &   &  $130$ &  $200$ &  $420$ &  $480$ & $450$ &  $22$ \\ Stat. & $^{+0.2}_{-0.3}$ &             $^{+0.3}_{-0.4}$ &             $^{+0.7}_{-0.9}$ &   &             $^{+20}_{-10}$ &   $^{+20}_{-20}$ &   $^{+10}_{-10}$ &             $^{+10}_{-10}$ &             $^{+10}_{-10}$ & \ldots &               $^{+0.4}_{-0.4}$ &             $^{+0.3}_{-0.3}$ & $^{+0.1}_{-0.1}$ &             $^{+0.3}_{-0.3}$ &   &             $^{+20}_{-10}$ &             $^{+20}_{-20}$ &             $^{+10}_{-10}$ &             $^{+10}_{-10}$ &             $^{+10}_{-10}$ &               $^{+2}_{-2}$ \\

\hline
\end{tabular}
\tablefoot{Results and statistical uncertainties (\textit{``Stat.'' row}) are given as median values and $1\sigma$ confidence limits, which are derived through a Monte Carlo simulation. The Galactic coordinate system is introduced in Fig.~\ref{fig:orbits}. Plane-crossing quantities are labeled by the subscript ``p'' and are based on all orbits that crossed the Galactic plane within twice the estimated stellar lifetime to account for the star's finite age (for HVS\,3 this limit is set to 200\,Myr, see text for details). The Galactic rest-frame velocity $\varv_{\mathrm{Grf}}=(\varv_x^2+\varv_y^2+\varv_z^2)^{1/2}$, the local Galactic escape velocity $\varv_{\mathrm{esc}}$, the Galactocentric radius $r=(x^2+y^2+z^2)^{1/2}$, the ejection velocity $\varv_{\mathrm{ej}}$ (defined as the Galactic rest-frame velocity relative to the rotating Galactic disk), and the flight time $\tau_{\mathrm{flight}}$ are listed in addition to Cartesian positions and velocities. The probability $P_{\mathrm{b}}$ is the fraction of Monte Carlo runs for which the star is bound to the Milky Way.}
\end{center}
\end{table*}
\begin{table*}
\begin{center}
\footnotesize
\setlength{\tabcolsep}{0.09cm}
\renewcommand{\arraystretch}{1.3}
\caption{\label{table:kinematic_parameters_II} Kinematic parameters of the program stars for Model~II of \citetads{2013A&A...549A.137I}.}
\begin{tabular}{lrrrrrrrrrrrrrrrrrrrrr}
\hline\hline
 & $x$ & $y$ & $z$ & & $\varv_x$ & $\varv_y$ & $\varv_z$ & $\varv_{\mathrm{Grf}}$ & $\varv_{\mathrm{Grf}}-\varv_{\mathrm{esc}}$ & $P_{\mathrm{b}}$ & $x_{\mathrm{p}}$ & $y_{\mathrm{p}}$ & $z_{\mathrm{p}}$ & $r_{\mathrm{p}}$ & & $\varv_{x\mathrm{,p}}$ & $\varv_{y\mathrm{,p}}$ & $\varv_{z\mathrm{,p}}$ & $\varv_{\mathrm{Grf,p}}$ &  $\varv_{\mathrm{ej,p}}$ & $\tau_{\mathrm{flight,p}}$ \\
\cline{2-4} \cline{6-10} \cline{12-15} \cline{17-21}
& \multicolumn{3}{c}{(kpc)} & & \multicolumn{5}{c}{$(\mathrm{km\,s^{-1}})$} & (\%) & \multicolumn{4}{c}{(kpc)} & & \multicolumn{5}{c}{$(\mathrm{km\,s^{-1}})$} & (Myr) \\
\hline \hline
 HVS\,1 (H) & $-65.9$ & $-62.4$ &  $51.6$ &   & $-420$ & $-310$ &  $440$ &  $700$ & $380$ &   $0$ & $-16.1$ & $-24.3$ &  $0.0$ &  $34.4$ &   & $-470$ & $-370$ &  $470$ &  $760$ & $740$ & $112$ \\ Stat. & $^{+5.4}_{-8.9}$ &             $^{+5.8}_{-9.6}$ &             $^{+8.0}_{-4.8}$ &   &           $^{+110}_{-100}$ &   $^{+80}_{-90}$ & $^{+110}_{-110}$ &             $^{+40}_{-30}$ &             $^{+40}_{-30}$ & \ldots &             $^{+22.9}_{-17.6}$ &           $^{+18.5}_{-14.2}$ & $^{+0.1}_{-0.1}$ &           $^{+16.8}_{-16.2}$ &   & $^{+110}_{-\phantom{0}70}$ & $^{+100}_{-\phantom{0}90}$ &             $^{+90}_{-60}$ &             $^{+40}_{-20}$ &             $^{+90}_{-90}$ &             $^{+35}_{-22}$ \\
     HVS\,3 & $-14.1$ & $-46.7$ & $-40.8$ &   & $-670$ & $-310$ & $-360$ &  $820$ & $450$ &   $0$ &  $56.3$ &  $-9.9$ &  $0.0$ &  $57.4$ &   & $-620$ & $-370$ & $-400$ &  $820$ & $730$ & $105$ \\ Stat. & $^{+0.8}_{-0.8}$ &             $^{+5.8}_{-5.9}$ &             $^{+5.1}_{-5.1}$ &   & $^{+100}_{-\phantom{0}90}$ &   $^{+30}_{-30}$ &   $^{+40}_{-30}$ &             $^{+80}_{-70}$ &             $^{+80}_{-80}$ & \ldots &             $^{+22.5}_{-17.8}$ &             $^{+4.7}_{-4.2}$ & $^{+0.1}_{-0.1}$ &           $^{+22.1}_{-17.3}$ &   & $^{+\phantom{0}90}_{-100}$ &             $^{+30}_{-20}$ &             $^{+40}_{-30}$ &             $^{+70}_{-50}$ &             $^{+80}_{-70}$ &             $^{+20}_{-18}$ \\
     HVS\,4 & $-64.2$ & $-14.7$ &  $52.9$ &   & $-360$ & $-320$ &  $360$ &  $640$ & $290$ &   $0$ & $-13.4$ &  $26.4$ &  $0.0$ &  $46.4$ &   & $-420$ & $-280$ &  $440$ &  $670$ & $820$ & $130$ \\ Stat. & $^{+5.2}_{-6.1}$ &             $^{+1.4}_{-1.6}$ &             $^{+5.9}_{-4.8}$ &   &           $^{+190}_{-180}$ & $^{+190}_{-190}$ & $^{+160}_{-170}$ & $^{+110}_{-\phantom{0}70}$ & $^{+120}_{-\phantom{0}60}$ & \ldots &             $^{+47.3}_{-32.4}$ &           $^{+23.2}_{-18.6}$ & $^{+0.1}_{-0.1}$ &           $^{+27.1}_{-22.7}$ &   &           $^{+210}_{-100}$ &           $^{+160}_{-210}$ &           $^{+110}_{-110}$ &             $^{+90}_{-30}$ & $^{+\phantom{0}80}_{-120}$ &             $^{+56}_{-35}$ \\
     HVS\,5 & $-28.6$ &  $13.5$ &  $19.5$ &   & $-400$ &  $330$ &  $400$ &  $650$ & $230$ &   $0$ &  $-8.6$ &  $-2.3$ &  $0.0$ &   $9.1$ &   & $-500$ &  $340$ &  $450$ &  $750$ & $630$ &  $46$ \\ Stat. & $^{+1.6}_{-2.1}$ &             $^{+1.5}_{-1.1}$ &             $^{+2.0}_{-1.6}$ &   &             $^{+30}_{-30}$ &   $^{+40}_{-40}$ &   $^{+30}_{-30}$ &             $^{+10}_{-10}$ &             $^{+10}_{-10}$ & \ldots &               $^{+1.9}_{-1.8}$ &             $^{+2.2}_{-2.1}$ & $^{+0.1}_{-0.1}$ &             $^{+1.7}_{-1.7}$ &   &             $^{+40}_{-40}$ &             $^{+40}_{-30}$ &             $^{+20}_{-20}$ &             $^{+20}_{-10}$ &             $^{+60}_{-40}$ &               $^{+5}_{-5}$ \\
     HVS\,6 & $-21.6$ & $-26.1$ &  $49.7$ &   & $-160$ & $-170$ &  $460$ &  $550$ & $170$ &   $0$ &  $-3.9$ &  $-6.0$ &  $0.0$ &  $19.5$ &   & $-220$ & $-250$ &  $570$ &  $650$ & $680$ &  $96$ \\ Stat. & $^{+1.7}_{-1.6}$ &             $^{+3.3}_{-3.0}$ &             $^{+5.8}_{-6.2}$ &   &           $^{+170}_{-170}$ & $^{+130}_{-140}$ &   $^{+90}_{-90}$ &             $^{+60}_{-30}$ &             $^{+60}_{-30}$ & \ldots &             $^{+18.2}_{-16.3}$ &           $^{+15.7}_{-13.2}$ & $^{+0.1}_{-0.1}$ &           $^{+14.0}_{-10.3}$ &   & $^{+170}_{-\phantom{0}80}$ & $^{+140}_{-\phantom{0}60}$ &             $^{+40}_{-70}$ &             $^{+40}_{-40}$ &             $^{+80}_{-80}$ &             $^{+22}_{-17}$ \\
     HVS\,7 & $-11.1$ & $-25.4$ &  $40.8$ &   & $-200$ &   $-0$ &  $450$ &  $500$ & $100$ &   $0$ &   $6.3$ & $-22.1$ &  $0.0$ &  $24.2$ &   & $-200$ &  $-90$ &  $510$ &  $560$ & $520$ &  $83$ \\ Stat. & $^{+0.2}_{-0.3}$ &             $^{+2.0}_{-2.4}$ &             $^{+3.7}_{-3.2}$ &   & $^{+\phantom{0}90}_{-100}$ &   $^{+50}_{-60}$ &   $^{+40}_{-30}$ &             $^{+50}_{-40}$ &             $^{+50}_{-40}$ & \ldots &               $^{+8.4}_{-7.2}$ &             $^{+4.7}_{-5.0}$ & $^{+0.1}_{-0.1}$ &             $^{+5.4}_{-4.8}$ &   &             $^{+70}_{-80}$ &             $^{+60}_{-60}$ &             $^{+30}_{-20}$ &             $^{+30}_{-20}$ &             $^{+30}_{-20}$ &               $^{+9}_{-8}$ \\
     HVS\,8 & $-30.3$ & $-13.5$ &  $26.9$ &   & $-420$ &   $80$ &  $260$ &  $500$ &  $90$ &   $0$ &   $9.4$ & $-16.5$ &  $0.0$ &  $19.9$ &   & $-450$ &  $-40$ &  $340$ &  $570$ & $440$ &  $88$ \\ Stat. & $^{+2.2}_{-2.6}$ &             $^{+1.4}_{-1.7}$ &             $^{+3.2}_{-2.6}$ &   &             $^{+70}_{-60}$ &   $^{+60}_{-70}$ &   $^{+50}_{-50}$ &             $^{+50}_{-40}$ &             $^{+50}_{-50}$ & \ldots &   $^{+11.5}_{-\phantom{0}8.1}$ &             $^{+5.3}_{-6.1}$ & $^{+0.1}_{-0.1}$ & $^{+10.1}_{-\phantom{0}6.3}$ &   &             $^{+40}_{-40}$ &             $^{+80}_{-70}$ &             $^{+40}_{-40}$ &             $^{+20}_{-20}$ &             $^{+50}_{-20}$ &             $^{+18}_{-14}$ \\
     HVS\,9 & $-28.8$ & $-43.0$ &  $46.5$ &   &  $-40$ & $-170$ &  $480$ &  $580$ & $210$ &   $0$ & $-22.8$ & $-24.2$ &  $0.0$ &  $42.0$ &   &  $-90$ & $-230$ &  $510$ &  $620$ & $690$ &  $90$ \\ Stat. & $^{+2.2}_{-1.9}$ &             $^{+4.6}_{-4.0}$ &             $^{+4.4}_{-4.9}$ &   &           $^{+250}_{-250}$ & $^{+170}_{-170}$ & $^{+150}_{-150}$ & $^{+130}_{-\phantom{0}90}$ & $^{+140}_{-\phantom{0}90}$ & \ldots &             $^{+25.2}_{-22.3}$ &           $^{+24.8}_{-17.0}$ & $^{+0.1}_{-0.1}$ &           $^{+16.3}_{-16.9}$ &   &           $^{+250}_{-210}$ &           $^{+190}_{-150}$ &           $^{+140}_{-100}$ & $^{+110}_{-\phantom{0}50}$ &           $^{+100}_{-120}$ &             $^{+35}_{-21}$ \\
HVS\,10 (H) & $-13.0$ & $-12.6$ &  $52.5$ &   & $-200$ & $-110$ &  $380$ &  $450$ &  $70$ &   $0$ &   $9.6$ &   $1.4$ &  $0.0$ &  $15.1$ &   & $-160$ & $-130$ &  $540$ &  $580$ & $590$ & $115$ \\ Stat. & $^{+0.5}_{-0.6}$ &             $^{+1.3}_{-1.4}$ &             $^{+6.1}_{-5.3}$ &   &           $^{+110}_{-100}$ & $^{+110}_{-110}$ &   $^{+30}_{-30}$ &             $^{+60}_{-30}$ &             $^{+60}_{-40}$ & \ldots &   $^{+13.1}_{-\phantom{0}9.6}$ &           $^{+12.8}_{-10.5}$ & $^{+0.1}_{-0.1}$ & $^{+13.2}_{-\phantom{0}8.4}$ &   &             $^{+50}_{-80}$ &             $^{+60}_{-50}$ &             $^{+60}_{-60}$ &             $^{+50}_{-30}$ &             $^{+70}_{-40}$ &             $^{+18}_{-14}$ \\
HVS\,12 (H) & $-20.6$ & $-29.0$ &  $41.0$ &   & $-240$ &  $-20$ &  $430$ &  $500$ & $110$ &   $0$ &   $1.8$ & $-23.9$ &  $0.0$ &  $25.7$ &   & $-250$ & $-110$ &  $480$ &  $560$ & $510$ &  $89$ \\ Stat. & $^{+1.4}_{-2.2}$ &             $^{+3.4}_{-5.1}$ &             $^{+7.2}_{-4.9}$ &   &             $^{+90}_{-90}$ &   $^{+80}_{-80}$ &   $^{+50}_{-60}$ &             $^{+60}_{-50}$ &             $^{+60}_{-50}$ & \ldots &   $^{+10.5}_{-\phantom{0}8.0}$ &             $^{+8.1}_{-8.7}$ & $^{+0.1}_{-0.1}$ &             $^{+9.1}_{-7.8}$ &   &             $^{+70}_{-80}$ & $^{+100}_{-\phantom{0}90}$ &             $^{+40}_{-40}$ &             $^{+40}_{-20}$ &             $^{+40}_{-40}$ &             $^{+19}_{-15}$ \\
HVS\,13 (H) & $-27.4$ & $-57.3$ &  $73.6$ &   & $-550$ &  $-50$ &  $360$ &  $690$ & $360$ &   $0$ &  $69.8$ & $-42.6$ &  $0.0$ &  $90.4$ &   & $-510$ & $-100$ &  $420$ &  $680$ & $620$ & $180$ \\ Stat. & $^{+2.6}_{-3.6}$ & $^{+\phantom{0}7.7}_{-10.8}$ & $^{+13.8}_{-\phantom{0}9.9}$ &   &           $^{+190}_{-200}$ & $^{+160}_{-160}$ & $^{+130}_{-130}$ &           $^{+170}_{-140}$ &           $^{+180}_{-150}$ & \ldots &             $^{+65.8}_{-39.1}$ &           $^{+37.5}_{-31.4}$ & $^{+0.1}_{-0.1}$ &           $^{+58.7}_{-36.3}$ &   &           $^{+180}_{-200}$ &           $^{+170}_{-150}$ &           $^{+100}_{-100}$ &           $^{+170}_{-110}$ &           $^{+170}_{-110}$ &             $^{+74}_{-45}$ \\
       B434 & $-16.0$ & $-22.4$ &  $32.9$ &   &  $120$ & $-280$ &  $220$ &  $380$ & $-30$ &  $72$ & $-24.7$ &  $13.6$ &  $0.0$ &  $29.7$ &   &   $-0$ & $-280$ &  $310$ &  $430$ & $580$ & $119$ \\ Stat. & $^{+0.7}_{-0.9}$ &             $^{+2.1}_{-2.6}$ &             $^{+3.9}_{-3.0}$ &   &             $^{+80}_{-70}$ &   $^{+60}_{-60}$ &   $^{+40}_{-50}$ &             $^{+60}_{-40}$ &             $^{+60}_{-40}$ & \ldots &   $^{+\phantom{0}9.1}_{-10.9}$ & $^{+13.6}_{-\phantom{0}8.9}$ & $^{+0.1}_{-0.1}$ & $^{+13.7}_{-\phantom{0}9.6}$ &   &             $^{+90}_{-90}$ &             $^{+30}_{-50}$ &             $^{+40}_{-40}$ &             $^{+20}_{-10}$ &             $^{+20}_{-20}$ &             $^{+27}_{-19}$ \\
       B485 & $-26.7$ &  $-5.9$ &  $27.2$ &   & $-330$ &  $140$ &  $270$ &  $450$ &  $30$ &   $2$ &   $4.5$ & $-14.9$ &  $0.0$ &  $15.7$ &   & $-380$ &   $20$ &  $370$ &  $540$ & $420$ &  $84$ \\ Stat. & $^{+1.0}_{-2.1}$ &             $^{+0.3}_{-0.7}$ &             $^{+3.0}_{-1.5}$ &   &             $^{+30}_{-30}$ &   $^{+20}_{-30}$ &   $^{+20}_{-20}$ &             $^{+20}_{-20}$ &             $^{+20}_{-20}$ & \ldots &               $^{+3.7}_{-2.8}$ &             $^{+1.7}_{-2.1}$ & $^{+0.1}_{-0.1}$ &             $^{+2.9}_{-2.1}$ &   &             $^{+10}_{-20}$ &             $^{+30}_{-30}$ &             $^{+20}_{-10}$ &             $^{+10}_{-10}$ &             $^{+10}_{-10}$ &   $^{+11}_{-\phantom{0}7}$ \\
       B711 &   $3.7$ &   $0.5$ &  $25.8$ &   &  $390$ &   $50$ &  $130$ &  $420$ & $-40$ &  $88$ & $-37.0$ &  $-4.4$ &  $0.0$ &  $37.4$ &   &  $270$ &   $30$ &  $260$ &  $380$ & $430$ & $116$ \\ Stat. & $^{+1.4}_{-0.9}$ &             $^{+0.1}_{-0.1}$ &             $^{+2.8}_{-2.0}$ &   &             $^{+40}_{-40}$ &   $^{+30}_{-40}$ &   $^{+20}_{-20}$ &             $^{+30}_{-30}$ &             $^{+40}_{-40}$ & \ldots &   $^{+\phantom{0}7.4}_{-11.2}$ &             $^{+3.3}_{-2.9}$ & $^{+0.1}_{-0.1}$ & $^{+11.1}_{-\phantom{0}7.3}$ &   &             $^{+50}_{-40}$ &             $^{+30}_{-20}$ &             $^{+30}_{-30}$ &             $^{+20}_{-10}$ &             $^{+20}_{-10}$ &             $^{+22}_{-16}$ \\
   B711 (H) &   $3.7$ &   $0.5$ &  $25.8$ &   &  $-90$ &  $330$ &  $350$ &  $510$ &  $50$ &  $32$ &   $7.9$ & $-18.9$ &  $0.0$ &  $22.0$ &   &  $-40$ &  $260$ &  $460$ &  $530$ & $600$ &  $62$ \\ Stat. & $^{+1.4}_{-1.0}$ &             $^{+0.1}_{-0.1}$ &             $^{+2.8}_{-2.1}$ &   &           $^{+130}_{-120}$ & $^{+140}_{-130}$ &   $^{+60}_{-60}$ &           $^{+120}_{-110}$ &           $^{+130}_{-110}$ & \ldots &               $^{+7.2}_{-7.3}$ &             $^{+8.4}_{-9.9}$ & $^{+0.1}_{-0.1}$ &             $^{+9.4}_{-8.1}$ &   &           $^{+100}_{-120}$ &           $^{+140}_{-140}$ &             $^{+50}_{-60}$ & $^{+100}_{-\phantom{0}50}$ &             $^{+80}_{-50}$ &   $^{+11}_{-\phantom{0}8}$ \\
       B733 &  $-5.5$ &   $3.3$ &   $8.9$ &   &  $230$ &  $190$ &  $350$ &  $460$ & $-80$ & $100$ &  $-9.9$ &  $-1.2$ &  $0.0$ &  $10.0$ &   &  $140$ &  $200$ &  $420$ &  $480$ & $450$ &  $22$ \\ Stat. & $^{+0.2}_{-0.3}$ &             $^{+0.3}_{-0.4}$ &             $^{+0.7}_{-0.9}$ &   &             $^{+20}_{-10}$ &   $^{+20}_{-20}$ &   $^{+10}_{-10}$ &             $^{+10}_{-10}$ &             $^{+10}_{-10}$ & \ldots &               $^{+0.3}_{-0.4}$ &             $^{+0.3}_{-0.4}$ & $^{+0.1}_{-0.1}$ &             $^{+0.3}_{-0.3}$ &   &             $^{+10}_{-20}$ &             $^{+20}_{-20}$ &             $^{+10}_{-10}$ &             $^{+10}_{-10}$ &             $^{+10}_{-10}$ &               $^{+2}_{-2}$ \\

\hline
\end{tabular}
\tablefoot{Same as for Table~\ref{table:kinematic_parameters_I}.}
\end{center}
\end{table*}
\begin{table*}
\begin{center}
\footnotesize
\setlength{\tabcolsep}{0.09cm}
\renewcommand{\arraystretch}{1.3}
\caption{\label{table:kinematic_parameters_III} Kinematic parameters of the program stars for Model~III of \citetads{2013A&A...549A.137I}.}
\begin{tabular}{lrrrrrrrrrrrrrrrrrrrrr}
\hline\hline
 & $x$ & $y$ & $z$ & & $\varv_x$ & $\varv_y$ & $\varv_z$ & $\varv_{\mathrm{Grf}}$ & $\varv_{\mathrm{Grf}}-\varv_{\mathrm{esc}}$ & $P_{\mathrm{b}}$ & $x_{\mathrm{p}}$ & $y_{\mathrm{p}}$ & $z_{\mathrm{p}}$ & $r_{\mathrm{p}}$ & & $\varv_{x\mathrm{,p}}$ & $\varv_{y\mathrm{,p}}$ & $\varv_{z\mathrm{,p}}$ & $\varv_{\mathrm{Grf,p}}$ &  $\varv_{\mathrm{ej,p}}$ & $\tau_{\mathrm{flight,p}}$ \\
\cline{2-4} \cline{6-10} \cline{12-15} \cline{17-21}
& \multicolumn{3}{c}{(kpc)} & & \multicolumn{5}{c}{$(\mathrm{km\,s^{-1}})$} & (\%) & \multicolumn{4}{c}{(kpc)} & & \multicolumn{5}{c}{$(\mathrm{km\,s^{-1}})$} & (Myr) \\
\hline \hline
 HVS\,1 (H) & $-65.9$ & $-62.4$ &  $51.6$ &   & $-420$ & $-310$ &  $440$ &  $700$ &  $120$ &   $0$ & $-15.5$ & $-23.3$ &  $0.0$ &  $32.8$ &   & $-500$ & $-410$ &  $490$ &  $810$ & $790$ & $110$ \\ Stat. & $^{+5.4}_{-8.9}$ &             $^{+5.8}_{-9.6}$ &             $^{+8.0}_{-4.8}$ &   &           $^{+110}_{-100}$ &   $^{+80}_{-90}$ & $^{+110}_{-110}$ &             $^{+40}_{-30}$ &             $^{+40}_{-30}$ & \ldots &             $^{+21.9}_{-17.3}$ &           $^{+18.0}_{-14.0}$ & $^{+0.1}_{-0.1}$ &           $^{+16.7}_{-15.8}$ &   & $^{+110}_{-\phantom{0}70}$ & $^{+100}_{-\phantom{0}90}$ &             $^{+90}_{-60}$ &             $^{+40}_{-20}$ &           $^{+110}_{-100}$ &             $^{+31}_{-22}$ \\
     HVS\,3 & $-14.1$ & $-46.7$ & $-40.8$ &   & $-670$ & $-310$ & $-360$ &  $820$ &  $180$ &   $1$ &  $52.7$ &  $-9.2$ &  $0.0$ &  $53.7$ &   & $-600$ & $-400$ & $-420$ &  $830$ & $710$ & $101$ \\ Stat. & $^{+0.8}_{-0.8}$ &             $^{+5.8}_{-5.9}$ &             $^{+5.1}_{-5.1}$ &   & $^{+100}_{-\phantom{0}90}$ &   $^{+20}_{-30}$ &   $^{+40}_{-30}$ &             $^{+80}_{-70}$ &             $^{+90}_{-80}$ & \ldots &             $^{+20.6}_{-16.4}$ &             $^{+4.5}_{-4.0}$ & $^{+0.1}_{-0.1}$ &           $^{+20.3}_{-15.9}$ &   &             $^{+90}_{-90}$ &             $^{+20}_{-30}$ &             $^{+30}_{-30}$ &             $^{+60}_{-50}$ &             $^{+70}_{-70}$ &             $^{+18}_{-17}$ \\
     HVS\,4 & $-64.2$ & $-14.7$ &  $52.9$ &   & $-360$ & $-320$ &  $360$ &  $640$ &   $30$ &  $32$ & $-11.0$ &  $24.5$ &  $0.0$ &  $43.6$ &   & $-480$ & $-260$ &  $470$ &  $720$ & $910$ & $126$ \\ Stat. & $^{+5.2}_{-6.1}$ &             $^{+1.4}_{-1.6}$ &             $^{+5.9}_{-4.8}$ &   &           $^{+190}_{-180}$ & $^{+190}_{-190}$ & $^{+160}_{-170}$ & $^{+110}_{-\phantom{0}70}$ & $^{+120}_{-\phantom{0}60}$ & \ldots &             $^{+45.2}_{-32.8}$ &           $^{+21.5}_{-17.5}$ & $^{+0.1}_{-0.1}$ &           $^{+26.7}_{-21.5}$ &   & $^{+220}_{-\phantom{0}70}$ &           $^{+140}_{-210}$ & $^{+100}_{-\phantom{0}90}$ &             $^{+70}_{-40}$ & $^{+\phantom{0}60}_{-150}$ &             $^{+50}_{-33}$ \\
     HVS\,5 & $-28.6$ &  $13.5$ &  $19.5$ &   & $-400$ &  $330$ &  $400$ &  $650$ &  $-40$ & $100$ &  $-8.4$ &  $-2.2$ &  $0.0$ &   $8.9$ &   & $-520$ &  $350$ &  $460$ &  $770$ & $660$ &  $45$ \\ Stat. & $^{+1.6}_{-2.1}$ &             $^{+1.5}_{-1.1}$ &             $^{+2.0}_{-1.6}$ &   &             $^{+30}_{-30}$ &   $^{+40}_{-50}$ &   $^{+30}_{-30}$ &             $^{+10}_{-10}$ &             $^{+10}_{-10}$ & \ldots &               $^{+1.8}_{-1.7}$ &             $^{+2.1}_{-2.1}$ & $^{+0.1}_{-0.1}$ &             $^{+1.7}_{-1.7}$ &   &             $^{+40}_{-40}$ &             $^{+30}_{-30}$ &             $^{+20}_{-20}$ &             $^{+20}_{-10}$ &             $^{+50}_{-50}$ &               $^{+5}_{-4}$ \\
     HVS\,6 & $-21.6$ & $-26.1$ &  $49.7$ &   & $-160$ & $-170$ &  $460$ &  $550$ &  $-90$ &  $92$ &  $-3.6$ &  $-5.6$ &  $0.0$ &  $18.2$ &   & $-230$ & $-270$ &  $600$ &  $690$ & $720$ &  $92$ \\ Stat. & $^{+1.7}_{-1.6}$ &             $^{+3.3}_{-3.0}$ &             $^{+5.8}_{-6.2}$ &   &           $^{+170}_{-170}$ & $^{+130}_{-140}$ &   $^{+90}_{-90}$ &             $^{+60}_{-30}$ &             $^{+60}_{-40}$ & \ldots &             $^{+16.8}_{-15.5}$ &           $^{+14.6}_{-12.5}$ & $^{+0.1}_{-0.1}$ & $^{+13.2}_{-\phantom{0}9.6}$ &   & $^{+150}_{-\phantom{0}80}$ & $^{+130}_{-\phantom{0}60}$ &             $^{+40}_{-60}$ &             $^{+40}_{-40}$ & $^{+100}_{-\phantom{0}90}$ &             $^{+20}_{-15}$ \\
     HVS\,7 & $-11.1$ & $-25.4$ &  $40.8$ &   & $-200$ &   $-0$ &  $450$ &  $500$ & $-160$ & $100$ &   $6.0$ & $-20.9$ &  $0.0$ &  $22.8$ &   & $-200$ & $-130$ &  $540$ &  $600$ & $550$ &  $80$ \\ Stat. & $^{+0.2}_{-0.3}$ &             $^{+2.0}_{-2.4}$ &             $^{+3.7}_{-3.2}$ &   & $^{+\phantom{0}90}_{-100}$ &   $^{+50}_{-60}$ &   $^{+40}_{-30}$ &             $^{+50}_{-40}$ &             $^{+50}_{-40}$ & \ldots &               $^{+7.8}_{-6.9}$ &             $^{+4.6}_{-4.7}$ & $^{+0.1}_{-0.1}$ &             $^{+5.2}_{-4.6}$ &   &             $^{+60}_{-70}$ &             $^{+60}_{-60}$ &             $^{+20}_{-20}$ &             $^{+20}_{-20}$ &             $^{+30}_{-20}$ &               $^{+9}_{-8}$ \\
     HVS\,8 & $-30.3$ & $-13.5$ &  $26.9$ &   & $-420$ &   $70$ &  $260$ &  $500$ & $-180$ & $100$ &   $8.8$ & $-15.2$ &  $0.0$ &  $18.5$ &   & $-460$ &  $-70$ &  $370$ &  $600$ & $460$ &  $84$ \\ Stat. & $^{+2.2}_{-2.6}$ &             $^{+1.4}_{-1.7}$ &             $^{+3.2}_{-2.6}$ &   &             $^{+70}_{-60}$ &   $^{+70}_{-60}$ &   $^{+50}_{-50}$ &             $^{+50}_{-40}$ &             $^{+60}_{-40}$ & \ldots &   $^{+10.2}_{-\phantom{0}7.5}$ &             $^{+4.8}_{-5.6}$ & $^{+0.1}_{-0.1}$ &             $^{+8.8}_{-5.8}$ &   &             $^{+30}_{-40}$ &             $^{+70}_{-70}$ &             $^{+30}_{-40}$ &             $^{+20}_{-10}$ &             $^{+40}_{-20}$ &             $^{+15}_{-13}$ \\
     HVS\,9 & $-28.8$ & $-43.0$ &  $46.5$ &   &  $-40$ & $-170$ &  $480$ &  $580$ &  $-50$ &  $66$ & $-21.7$ & $-23.1$ &  $0.0$ &  $40.2$ &   & $-120$ & $-270$ &  $530$ &  $650$ & $740$ &  $88$ \\ Stat. & $^{+2.2}_{-1.9}$ &             $^{+4.6}_{-4.0}$ &             $^{+4.4}_{-4.9}$ &   &           $^{+250}_{-250}$ & $^{+170}_{-180}$ & $^{+150}_{-150}$ & $^{+130}_{-\phantom{0}90}$ & $^{+140}_{-\phantom{0}90}$ & \ldots &             $^{+24.0}_{-21.6}$ &           $^{+23.8}_{-16.9}$ & $^{+0.1}_{-0.1}$ &           $^{+16.2}_{-16.8}$ &   &           $^{+250}_{-200}$ &           $^{+190}_{-140}$ & $^{+130}_{-\phantom{0}90}$ &             $^{+90}_{-40}$ &           $^{+110}_{-140}$ &             $^{+31}_{-20}$ \\
HVS\,10 (H) & $-13.0$ & $-12.6$ &  $52.5$ &   & $-200$ & $-110$ &  $380$ &  $460$ & $-200$ & $100$ &   $8.8$ &   $1.3$ &  $0.0$ &  $13.8$ &   & $-160$ & $-140$ &  $580$ &  $620$ & $640$ & $108$ \\ Stat. & $^{+0.5}_{-0.6}$ &             $^{+1.3}_{-1.4}$ &             $^{+6.1}_{-5.3}$ &   &           $^{+110}_{-100}$ & $^{+110}_{-110}$ &   $^{+30}_{-30}$ &             $^{+50}_{-40}$ &             $^{+60}_{-30}$ & \ldots &   $^{+11.5}_{-\phantom{0}8.8}$ & $^{+11.5}_{-\phantom{0}9.5}$ & $^{+0.1}_{-0.1}$ & $^{+11.5}_{-\phantom{0}7.5}$ &   &             $^{+40}_{-60}$ &             $^{+50}_{-40}$ &             $^{+60}_{-40}$ &             $^{+50}_{-20}$ &             $^{+70}_{-40}$ &             $^{+16}_{-12}$ \\
HVS\,12 (H) & $-20.6$ & $-29.0$ &  $41.0$ &   & $-240$ &  $-20$ &  $430$ &  $500$ & $-160$ &  $99$ &   $1.7$ & $-22.5$ &  $0.0$ &  $24.2$ &   & $-270$ & $-150$ &  $510$ &  $600$ & $540$ &  $86$ \\ Stat. & $^{+1.4}_{-2.2}$ &             $^{+3.4}_{-5.1}$ &             $^{+7.2}_{-4.9}$ &   &             $^{+90}_{-90}$ &   $^{+70}_{-80}$ &   $^{+50}_{-60}$ &             $^{+60}_{-50}$ &             $^{+70}_{-50}$ & \ldots &               $^{+9.8}_{-7.6}$ &             $^{+7.8}_{-8.4}$ & $^{+0.1}_{-0.1}$ &             $^{+8.7}_{-7.5}$ &   &             $^{+70}_{-60}$ &             $^{+90}_{-90}$ &             $^{+40}_{-30}$ &             $^{+30}_{-20}$ &             $^{+40}_{-40}$ &             $^{+17}_{-14}$ \\
HVS\,13 (H) & $-27.4$ & $-57.3$ &  $73.6$ &   & $-550$ &  $-50$ &  $360$ &  $690$ &  $100$ &  $25$ &  $64.5$ & $-38.0$ &  $0.0$ &  $83.6$ &   & $-480$ & $-160$ &  $460$ &  $690$ & $610$ & $171$ \\ Stat. & $^{+2.6}_{-3.6}$ & $^{+\phantom{0}7.7}_{-10.8}$ & $^{+13.8}_{-\phantom{0}9.9}$ &   &           $^{+190}_{-200}$ & $^{+160}_{-160}$ & $^{+130}_{-130}$ &           $^{+170}_{-140}$ &           $^{+190}_{-150}$ & \ldots &             $^{+59.9}_{-35.8}$ &           $^{+36.8}_{-31.3}$ & $^{+0.1}_{-0.1}$ &           $^{+54.3}_{-34.3}$ &   &           $^{+170}_{-200}$ &           $^{+180}_{-150}$ & $^{+\phantom{0}90}_{-100}$ & $^{+150}_{-\phantom{0}70}$ & $^{+150}_{-\phantom{0}80}$ &             $^{+66}_{-40}$ \\
       B434 & $-16.0$ & $-22.4$ &  $32.9$ &   &  $120$ & $-280$ &  $220$ &  $380$ & $-300$ & $100$ & $-21.9$ &  $12.2$ &  $0.0$ &  $26.4$ &   &  $-50$ & $-290$ &  $350$ &  $460$ & $640$ & $110$ \\ Stat. & $^{+0.7}_{-0.9}$ &             $^{+2.1}_{-2.6}$ &             $^{+3.9}_{-3.0}$ &   &             $^{+80}_{-70}$ &   $^{+60}_{-60}$ &   $^{+40}_{-50}$ &             $^{+60}_{-40}$ &             $^{+60}_{-40}$ & \ldots &               $^{+8.1}_{-9.5}$ & $^{+11.2}_{-\phantom{0}7.9}$ & $^{+0.1}_{-0.1}$ & $^{+11.4}_{-\phantom{0}8.3}$ &   &             $^{+90}_{-80}$ &             $^{+30}_{-40}$ &             $^{+30}_{-30}$ &             $^{+20}_{-10}$ &             $^{+20}_{-10}$ &             $^{+22}_{-16}$ \\
       B485 & $-26.7$ &  $-5.9$ &  $27.2$ &   & $-330$ &  $140$ &  $270$ &  $450$ & $-240$ & $100$ &   $4.2$ & $-13.8$ &  $0.0$ &  $14.6$ &   & $-410$ &   $-0$ &  $400$ &  $570$ & $440$ &  $80$ \\ Stat. & $^{+1.0}_{-2.1}$ &             $^{+0.3}_{-0.7}$ &             $^{+3.0}_{-1.5}$ &   &             $^{+30}_{-30}$ &   $^{+20}_{-30}$ &   $^{+20}_{-20}$ &             $^{+20}_{-20}$ &             $^{+20}_{-20}$ & \ldots &               $^{+3.4}_{-2.6}$ &             $^{+1.5}_{-1.8}$ & $^{+0.1}_{-0.1}$ &             $^{+2.6}_{-1.9}$ &   &             $^{+10}_{-10}$ &             $^{+20}_{-30}$ &             $^{+10}_{-20}$ &             $^{+10}_{-10}$ &             $^{+20}_{-10}$ &   $^{+10}_{-\phantom{0}6}$ \\
       B711 &   $3.7$ &   $0.5$ &  $25.8$ &   &  $390$ &   $50$ &  $130$ &  $420$ & $-310$ & $100$ & $-32.2$ &  $-3.8$ &  $0.0$ &  $32.5$ &   &  $240$ &   $30$ &  $300$ &  $380$ & $460$ & $105$ \\ Stat. & $^{+1.4}_{-0.9}$ &             $^{+0.1}_{-0.1}$ &             $^{+2.8}_{-2.0}$ &   &             $^{+40}_{-40}$ &   $^{+30}_{-40}$ &   $^{+20}_{-20}$ &             $^{+30}_{-30}$ &             $^{+40}_{-30}$ & \ldots &               $^{+5.8}_{-8.3}$ &             $^{+2.9}_{-2.5}$ & $^{+0.1}_{-0.1}$ &             $^{+8.2}_{-5.7}$ &   &             $^{+30}_{-30}$ &             $^{+20}_{-20}$ &             $^{+20}_{-20}$ &             $^{+20}_{-10}$ &             $^{+10}_{-10}$ &             $^{+17}_{-13}$ \\
   B711 (H) &   $3.7$ &   $0.5$ &  $25.8$ &   &  $-90$ &  $330$ &  $350$ &  $510$ & $-210$ &  $96$ &   $7.6$ & $-18.0$ &  $0.0$ &  $20.9$ &   &  $-30$ &  $240$ &  $480$ &  $540$ & $620$ &  $60$ \\ Stat. & $^{+1.4}_{-1.0}$ &             $^{+0.1}_{-0.1}$ &             $^{+2.8}_{-2.1}$ &   &           $^{+130}_{-120}$ & $^{+140}_{-140}$ &   $^{+60}_{-60}$ &           $^{+120}_{-110}$ &           $^{+120}_{-110}$ & \ldots &               $^{+6.9}_{-7.0}$ &             $^{+8.0}_{-9.2}$ & $^{+0.1}_{-0.1}$ &             $^{+8.8}_{-7.7}$ &   & $^{+\phantom{0}90}_{-110}$ &           $^{+140}_{-130}$ &             $^{+50}_{-60}$ &             $^{+90}_{-50}$ &             $^{+80}_{-50}$ &   $^{+10}_{-\phantom{0}8}$ \\
       B733 &  $-5.5$ &   $3.3$ &   $8.9$ &   &  $230$ &  $190$ &  $350$ &  $460$ & $-330$ & $100$ &  $-9.9$ &  $-1.2$ &  $0.0$ &  $10.0$ &   &  $140$ &  $200$ &  $420$ &  $480$ & $450$ &  $22$ \\ Stat. & $^{+0.2}_{-0.3}$ &             $^{+0.3}_{-0.4}$ &             $^{+0.7}_{-0.9}$ &   &             $^{+20}_{-10}$ &   $^{+20}_{-20}$ &   $^{+10}_{-10}$ &             $^{+10}_{-10}$ &             $^{+10}_{-10}$ & \ldots &               $^{+0.3}_{-0.4}$ &             $^{+0.3}_{-0.3}$ & $^{+0.1}_{-0.1}$ &             $^{+0.3}_{-0.3}$ &   &             $^{+10}_{-20}$ &             $^{+20}_{-20}$ &             $^{+10}_{-20}$ &             $^{+10}_{-10}$ &             $^{+10}_{-10}$ &               $^{+2}_{-2}$ \\

\hline
\end{tabular}
\tablefoot{Same as for Table~\ref{table:kinematic_parameters_I}.}
\end{center}
\end{table*}
\begin{figure*}
\begin{center}
\includegraphics[width=0.33\textwidth]{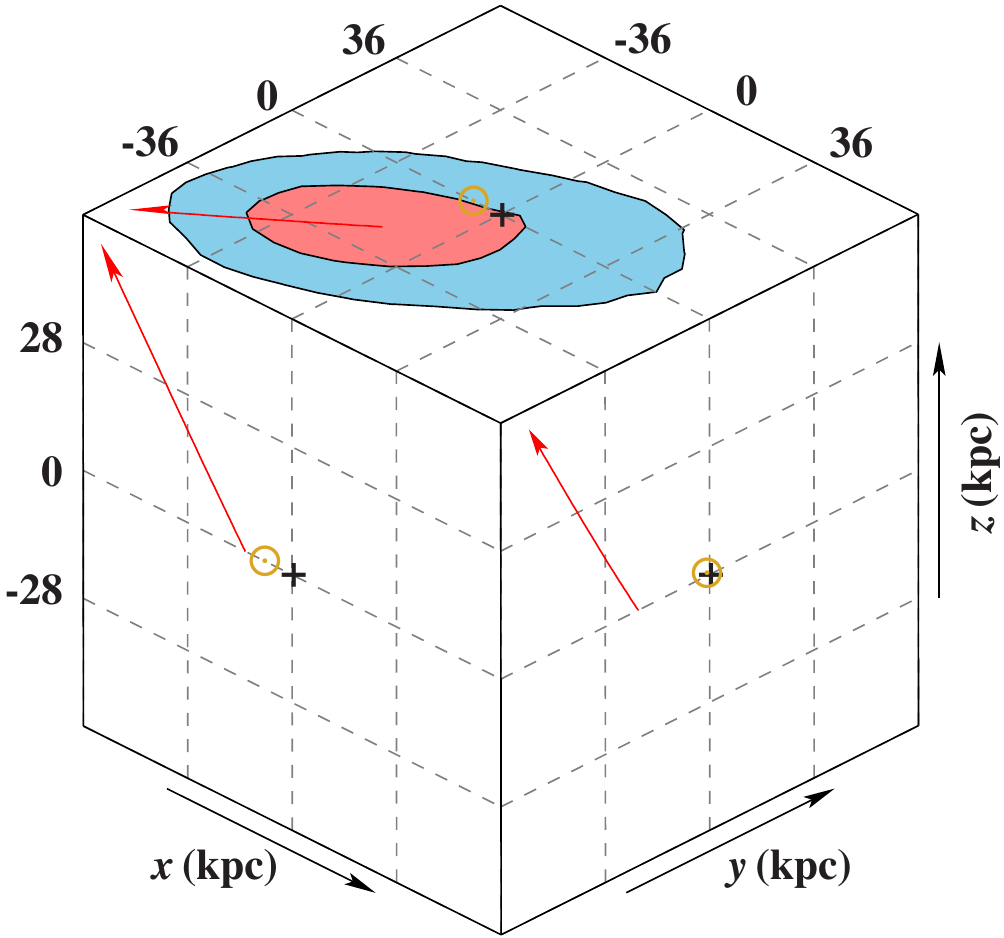}
\includegraphics[width=0.33\textwidth]{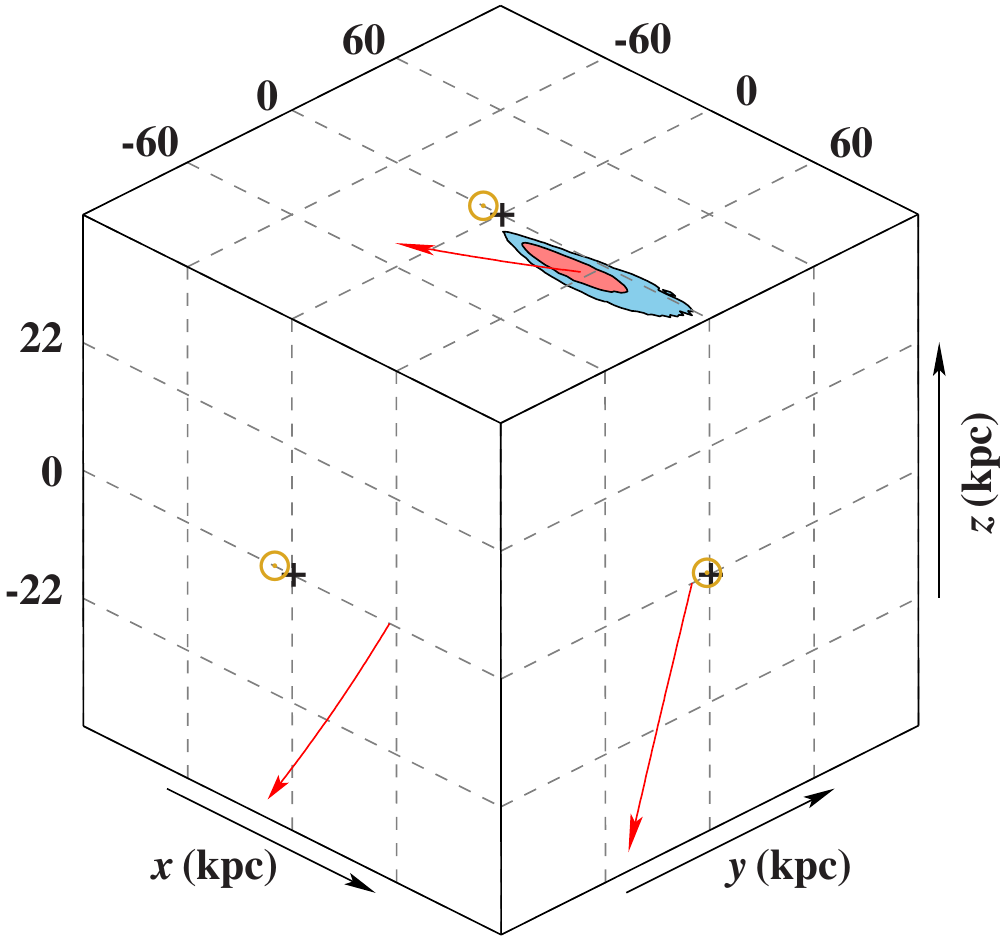}
\includegraphics[width=0.33\textwidth]{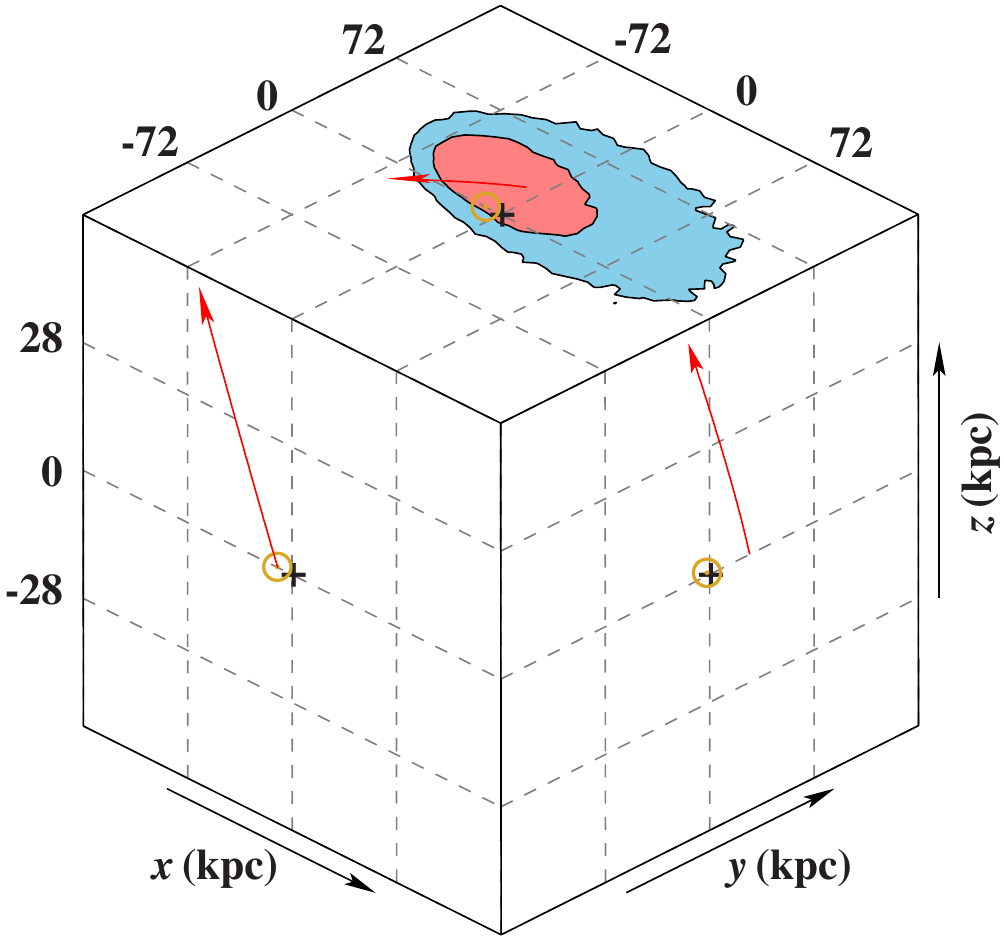}
\includegraphics[width=0.33\textwidth]{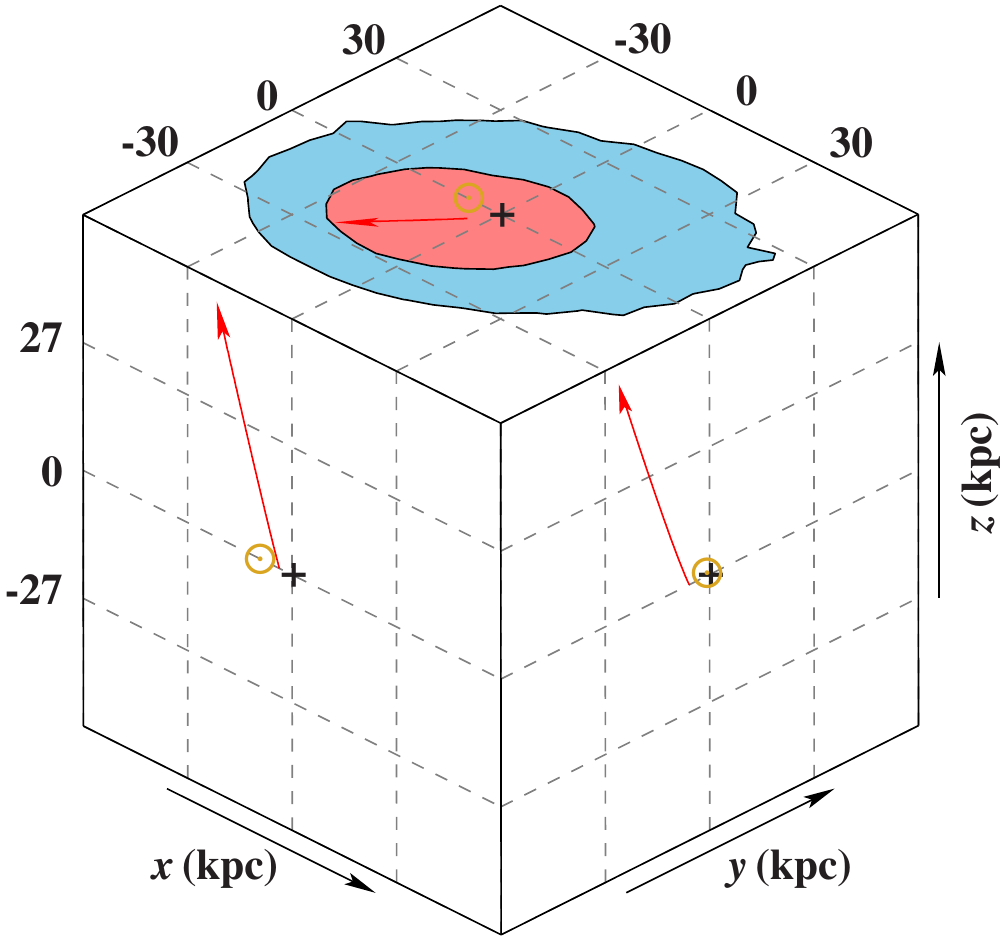}
\includegraphics[width=0.33\textwidth]{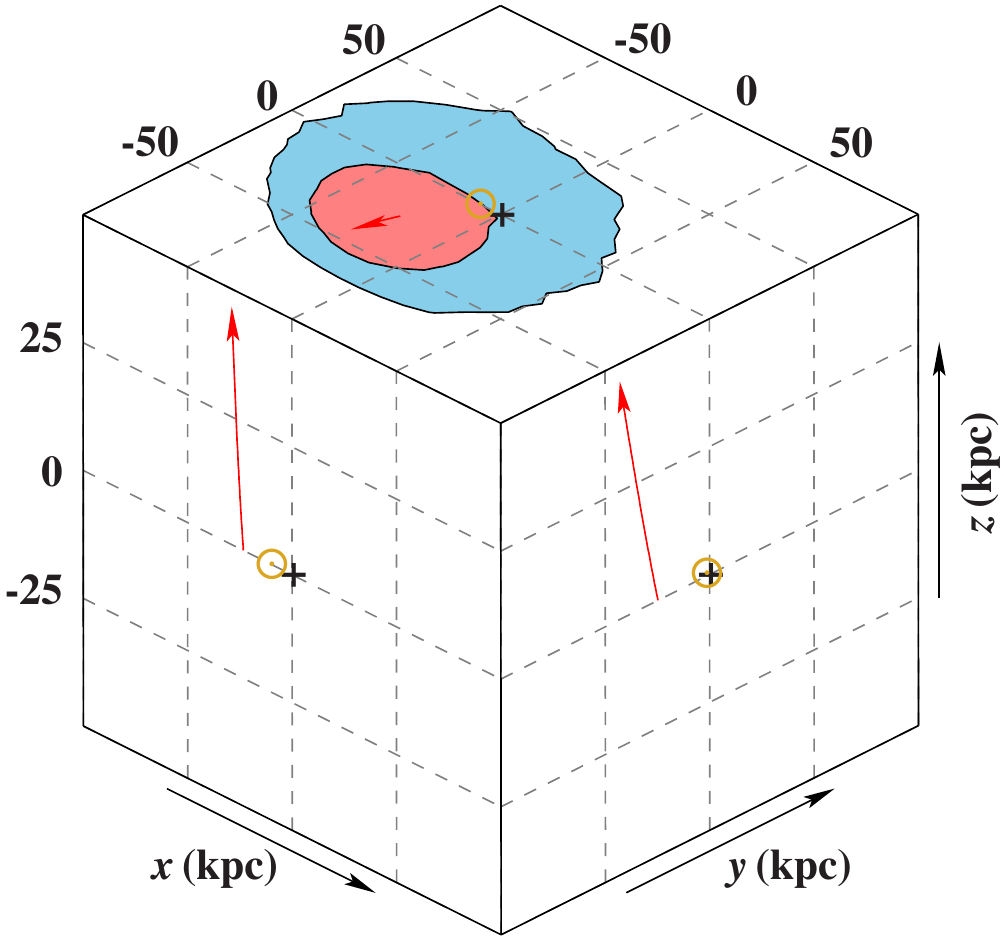}
\includegraphics[width=0.33\textwidth]{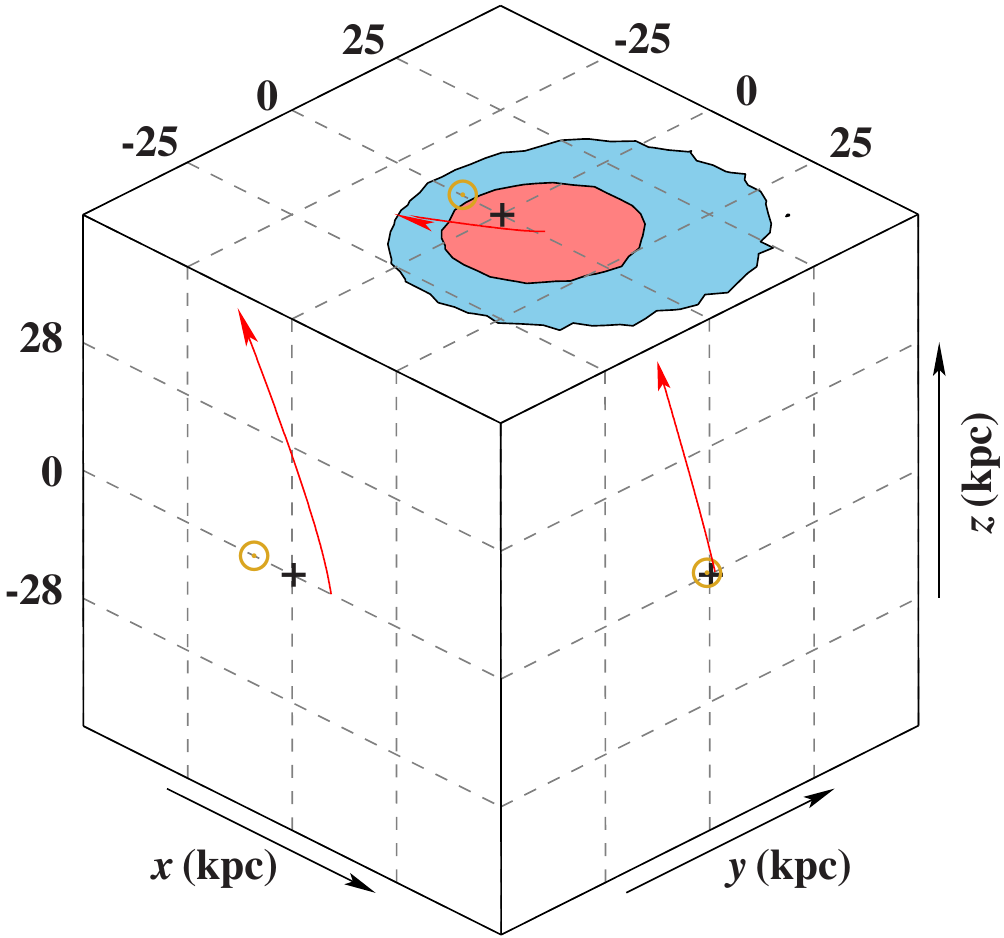}
\includegraphics[width=0.33\textwidth]{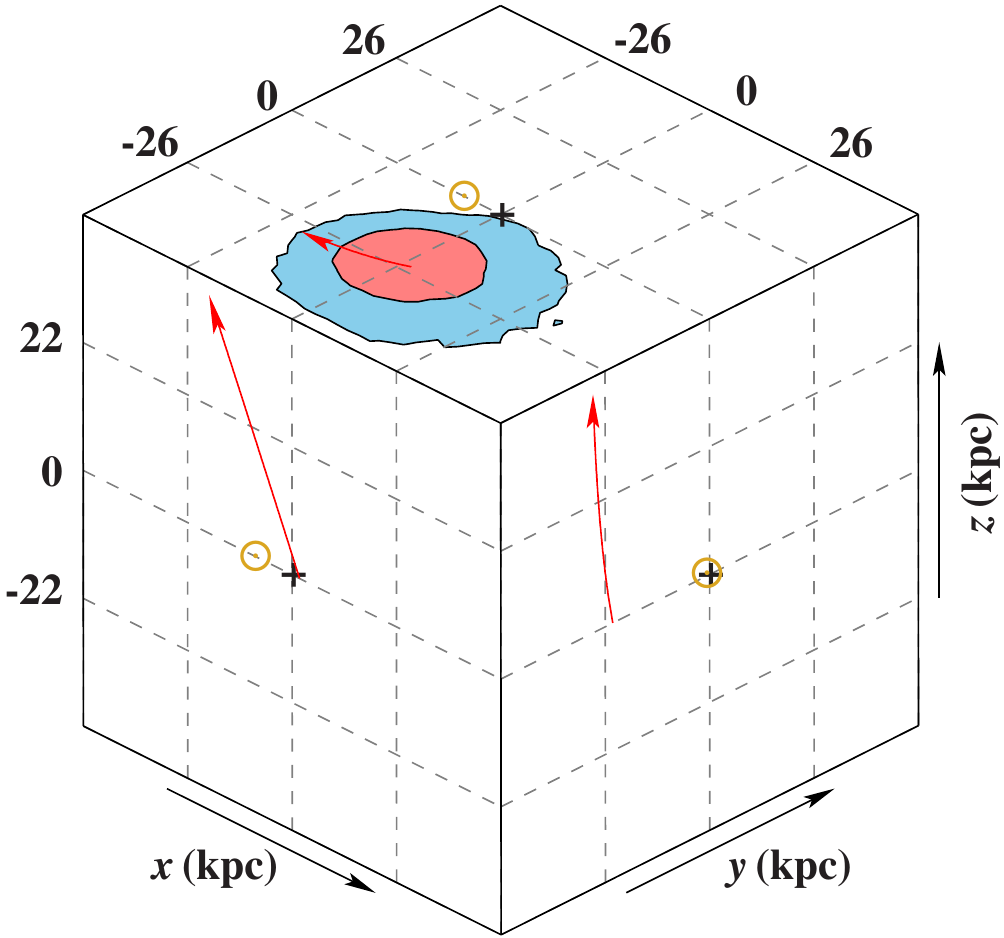}
\includegraphics[width=0.33\textwidth]{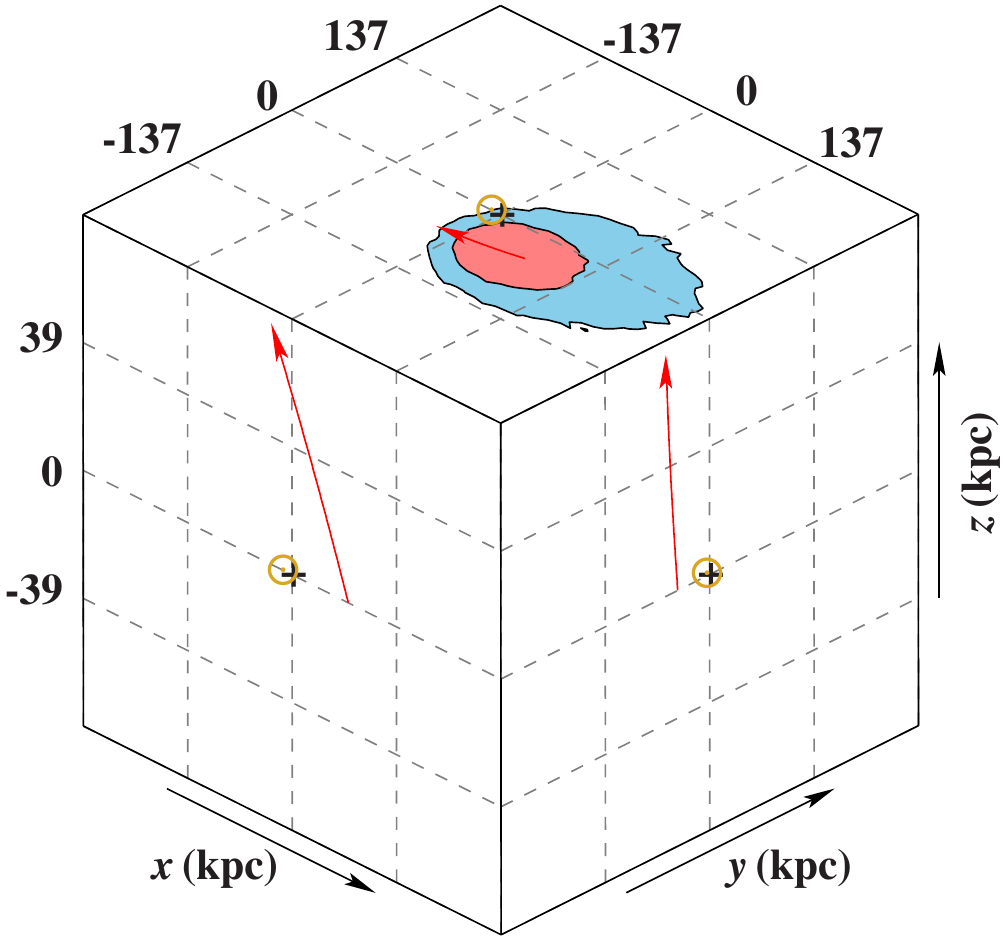}
\includegraphics[width=0.33\textwidth]{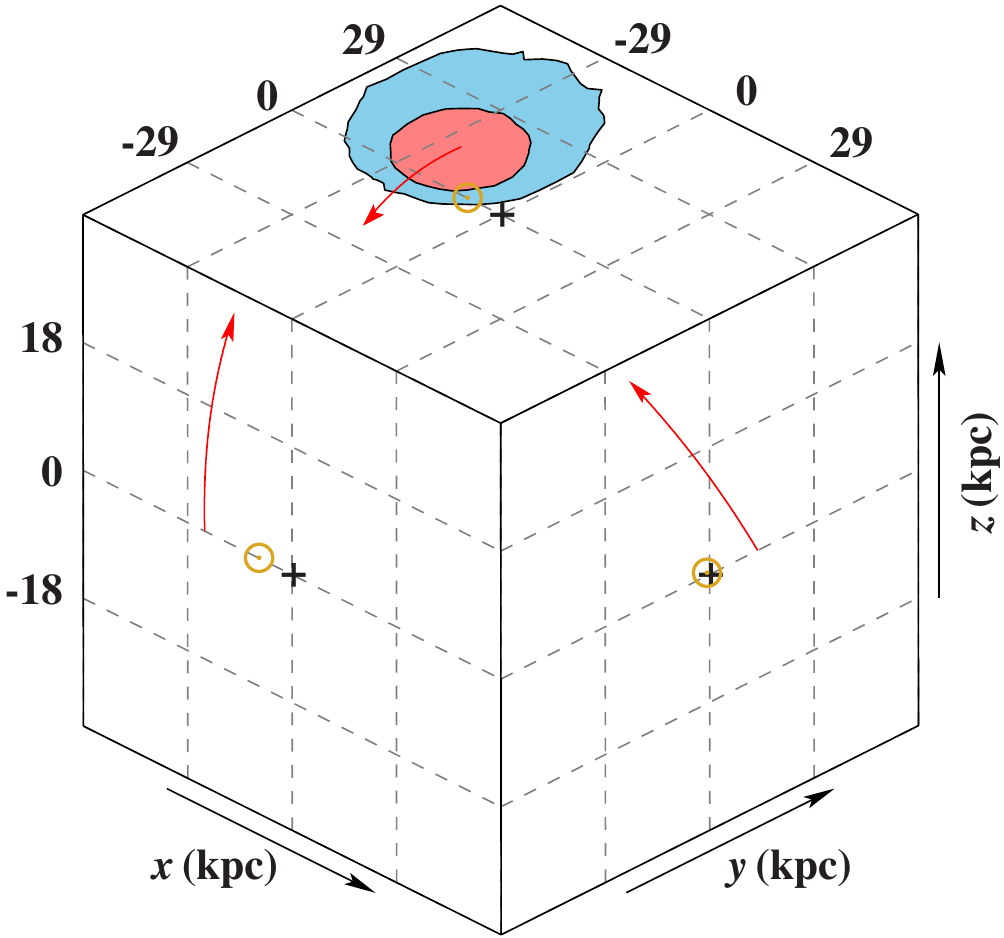}
\caption{Same as Fig.~\ref{fig:orbits}, but for HVS\,1 (H) (\textit{top left}), HVS\,3 (\textit{top center}), HVS\,4 (\textit{top right}), HVS\,6 (\textit{middle left}), HVS\,9 (\textit{middle center}), HVS\,10 (H) (\textit{middle right}), HVS\,12 (H) (\textit{bottom left}), HVS\,13 (H) (\textit{bottom center}), and B434 (\textit{bottom right}).}
\label{fig:orbits_appendix}
\end{center}
\end{figure*}
\end{appendix}
\end{document}